\documentclass{raa_twocolumn}
\usepackage{graphicx,times}             
\usepackage{natbib}
\usepackage{amssymb,amsmath}
\bibpunct{(}{)}{;}{a}{}{,}
\usepackage[pagebackref=true]{hyperref}
\usepackage{multicol}
\usepackage{multirow}
\usepackage{subcaption}
\begin{document}


\title{The trigger and localization system of SVOM-GRM}

   \volnopage{Vol.0 (20xx) No.0, 000--000}      
   \setcounter{page}{1}          

   \author{Jiang He
      \inst{1,*}
   \and Jian-Chao Sun
      \inst{1,*}
    \and Yue Huang 
        \inst{1}
   \and Yong-Wei Dong
        \inst{1}
    \and Shi-Jie Zheng
        \inst{1}
    \and Xiao-Yun Zhao
        \inst{1}
    \and Min Gao
        \inst{1}
    \and Lu Li
        \inst{1}
    \and Jiang-Tao Liu
        \inst{1}
    \and Xin Liu
        \inst{1}
    \and Hao-Li Shi
        \inst{1}
    \and Li-Ming Song
        \inst{1,2}
    \and Wen-Jun Tan
        \inst{1,2}
    \and Bo-Bing Wu
        \inst{1,2}
    \and Chen-Wei Wang
        \inst{1,2}
    \and Jin Wang
        \inst{1}
    \and Jin-Zhou Wang
        \inst{1}
    \and Ping Wang
        \inst{1}
    \and Rui-Jie Wang
        \inst{1}
    \and Shao-lin Xiong
        \inst{1,2,*}
    \and Juan Zhang
        \inst{1}
    \and Li Zhang
        \inst{1}
    \and Shuang-Nan Zhang
        \inst{1,2}
}

   \institute{
   State Key Laboratory of Particle Astrophysics, Institute of High Energy Physics, Chinese Academy of Sciences,Beijing 100049, China; {\it hejiang@ihep.ac.cn, sunjc@ihep.ac.cn, xiongsl@ihep.ac.cn}\\
    \and University of Chinese Academy of Sciences, Chinese Academy of Sciences, Beijing 100049, China\\
\vs\no
   {\small Received 20xx month day; accepted 20xx month day}}

\abstract{The Space multi-band Variable Object Monitor (SVOM) is an astronomical satellite jointly developed by China and France, primarily focused on the detection of gamma-ray bursts (GRBs) and transient sources. The SVOM satellite was launched on 22nd June, 2024 with four payloads installed onboard. As one of payload, GRM comprises 3 gamma-ray detectors (each detector has an effective area of approximately 200~cm$^{2}$) with distinct pointing directions, enabling the temporal and spectral measurements as well as localization of GRBs in the energy range of 15-5000 keV. This article firstly introduces the on-board localization algorithm design for GRM and presents preliminary test results. Then, leveraging abundant ground-based computational resources, a joint fitting method for spectral and localization analysis using Monte Carlo Markov Chain (MCMC) is implemented. In contrast to the on-board localization algorithm, the on-ground MCMC method comprehensively considers the influence of spectral characteristics, thereby mitigating systematic biases. Finally, a systematic analysis based on this method is provided, highlighting the localization and spectral measurement capabilities of GRM. The preliminary localization analysis result for the on-board detected GRB 240629A 
by both GRM and Fermi/GBM shows that the localization result (error$\sim$4.14$^{\circ}$) of GRM is consistent with the Fermi/GBM result.
}

   \authorrunning{Jiang, He \& Jian-Chao Sun }            
   \titlerunning{GRM trigger and localization}  

   \maketitle
%
%
\section{Introduction}           
\label{sect:intro}

Since their discovery, gamma ray bursts (GRBs) have been studied for over five decades~\citep{klebesadel1973,gehrels2009}. Research on GRBs holds substantial importance in realms encompassing particle acceleration mechanisms, radiation mechanisms, and constraints on cosmological parameters~\citep{zhang2011,zhang2018}.

The current operational satellites, such as Swift, Fermi, INTEGRAL, Insight-HXMT and GECAM have generated a wealth of commendable results, including the discovery of the correlation between GRB 170817A and gravitational waves GW170817~\citep{abbott2017,li2018,goldstein2017}, the thermal components of GRBs~\citep{guiriec2011}, and the brightest GRB(GRB~221009A) of all time~\citep{zhenghua2023,lesage2023,zhang2025,zhangyanqiu2024,zhangyanqiu2024b}. The Sino-French collaborative multi-band astronomical observation satellite, SVOM, features onboard triggering and localization capabilities. SVOM can transmit trigger and localization information to ground in real-time by using the VHF~\citep{sebastien2018} or Beidou short messages, guiding ground-based telescopes for follow-up observations. The collaborative observation efforts of SVOM with other instruments such as IceCube~\citep{achterberg2006}, LIGO~\citep{aasi2015}, will herald an exciting era of multi-messenger observations of GRBs.

The capability to trigger and locate GRBs represents a key scientific requirement for SVOM. SVOM comprises four instruments, among which Eclairs and GRM possess on-orbit triggering and localization abilities, operating within the 4-100~keV and 15-5000~keV energy ranges respectively. This article primarily concentrates on the triggering and localization capabilities of GRM~\citep{zhaodonghua2013,hejiang2025}. Given the limited computing resources and the need for rapid trigger and localization results, the primary method for on-orbit localization involves traversing tables to calculate chi-square values, as detailed in Chapter 1. Compared with on-orbit localization, offline localization using scientific data has a certain time delay, but can achieve better results. In Chapter 3, a location method based on Bayesian probability is introduced, and the expected localization ability of GRM is demonstrated through simulation.
\section{An Introduction of GRM}
\begin{figure}[!htbp]
    \centering
    \includegraphics[width=\linewidth]{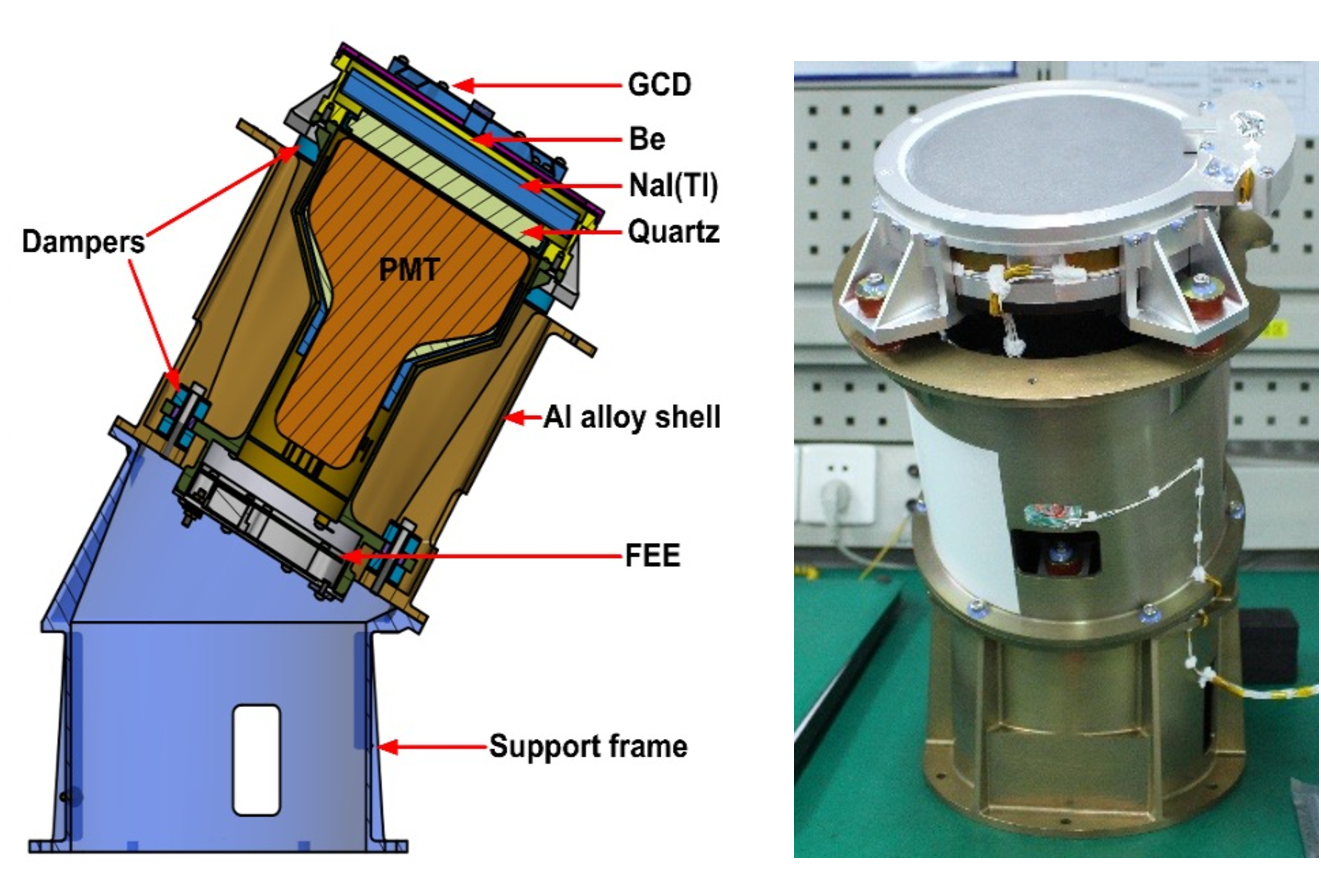}
    \caption{The schematic diagram of the GRD structure of the SVOM satellite.}
    \label{fig:svom_grd}
\end{figure}

The Gamma-Ray Monitor (GRM) consists of three Gamma-Ray Detectors (GRDs), one GRM Particle Monitor (GPM), and one GRM Electronic Box (GEB). The optical axes of the three GRDs are oriented at an angle of 30$^{\circ}$ relative to the +Z-axis direction, with their projections onto the X-Y plane separated by mutual azimuthal angles of 120$^{circ}$ to ensure full-sky coverage. Each GRD detector is equipped with a NaI(Tl) scintillator crystal, 16~cm in diameter and 1.5~cm in thickness, operating over a energy range of 15 keV to 5 MeV. The Figure~\ref{fig:svom_grd} is the GRD's structure, for further details on the GRM instrument’s design and performance specifications, refer to the corresponding references~\citep{Sun+etal+2026}.
 
\section{Introduction of On-Orbit Triggering and Localization}
\label{sect:on-board}
\subsection{Triggering and Localization Process}
\begin{figure}[!htbp]
    \centering
    \includegraphics[width=\linewidth]{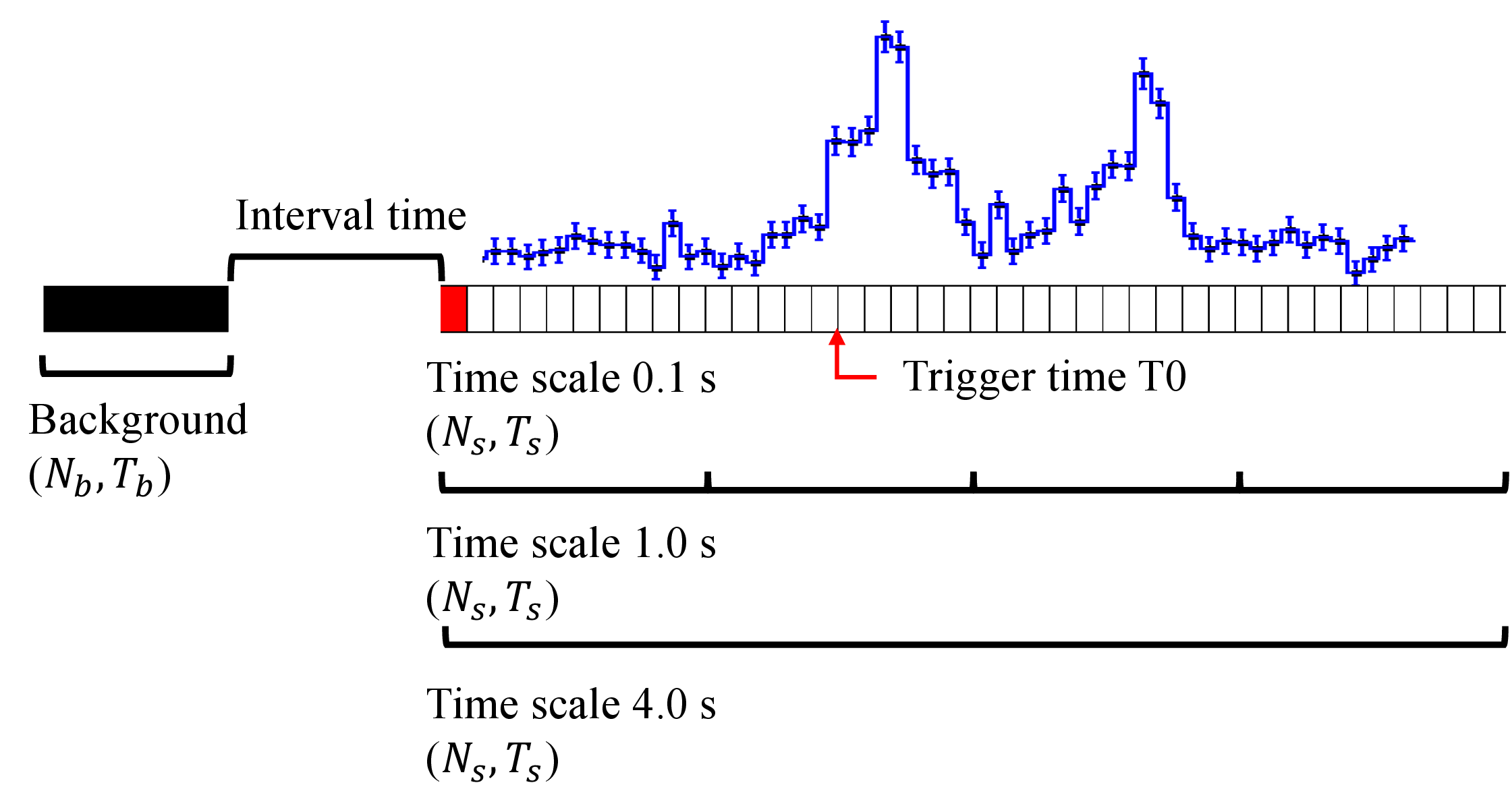}
    \caption{The schematic diagram of trigger calculation, where $N_\mathrm{s}$ and $N_\mathrm{b}$ correspond to the counts under trigger time scale $T_\mathrm{s}$ and background time scale $T_\mathrm{b}$ respectively. The $T_\mathrm{b}$ time scale and the interval time are adjustable.}
    \label{fig:tri_process}
\end{figure}

The GRM employs a count rate trigger algorithm operating across distinct energy bands and time windows. As shown in Figure~\ref{fig:tri_process}, the onboard software calculates the observed photon counts \(N_{\mathrm{s}}\) within a selected energy band (e.g., 15--50~keV) over a specified integration time \(T_{\mathrm{s}}\). Simultaneously, it estimates the background counts \(N_{\mathrm{b}}\) using the same energy band from a preceding interval \(T_{\mathrm{b}}\). The signal-to-noise ratio (snr) for each GRD is derived using equation:

\begin{equation}\label{eq:snr}
    \mathrm{snr}=\frac{N_\mathrm{s}-\frac{N_\mathrm{b}\times T_\mathrm{s}}{T_\mathrm{b}}}{\sqrt{\frac{N_\mathrm{b}\times T_\mathrm{s}}{T_\mathrm{b}}}},
\end{equation}

\noindent as defined in Eq.~\ref{eq:snr}. Trigger conditions are met only when the snr~values from at least two GRDs simultaneously exceed predefined thresholds. The algorithm incorporates three configurable time scales (0.1~s, 1~s, and 4~s) and four adjustable energy bands: 15--50~keV, 50--300~keV, 300--1000~keV, and 1000--5000~keV. These parameters can be updated in-flight to optimize sensitivity for different astrophysical transients while minimizing false triggers caused by background fluctuations.

\begin{figure}[!htbp]
    \centering
    \includegraphics[width=\linewidth]{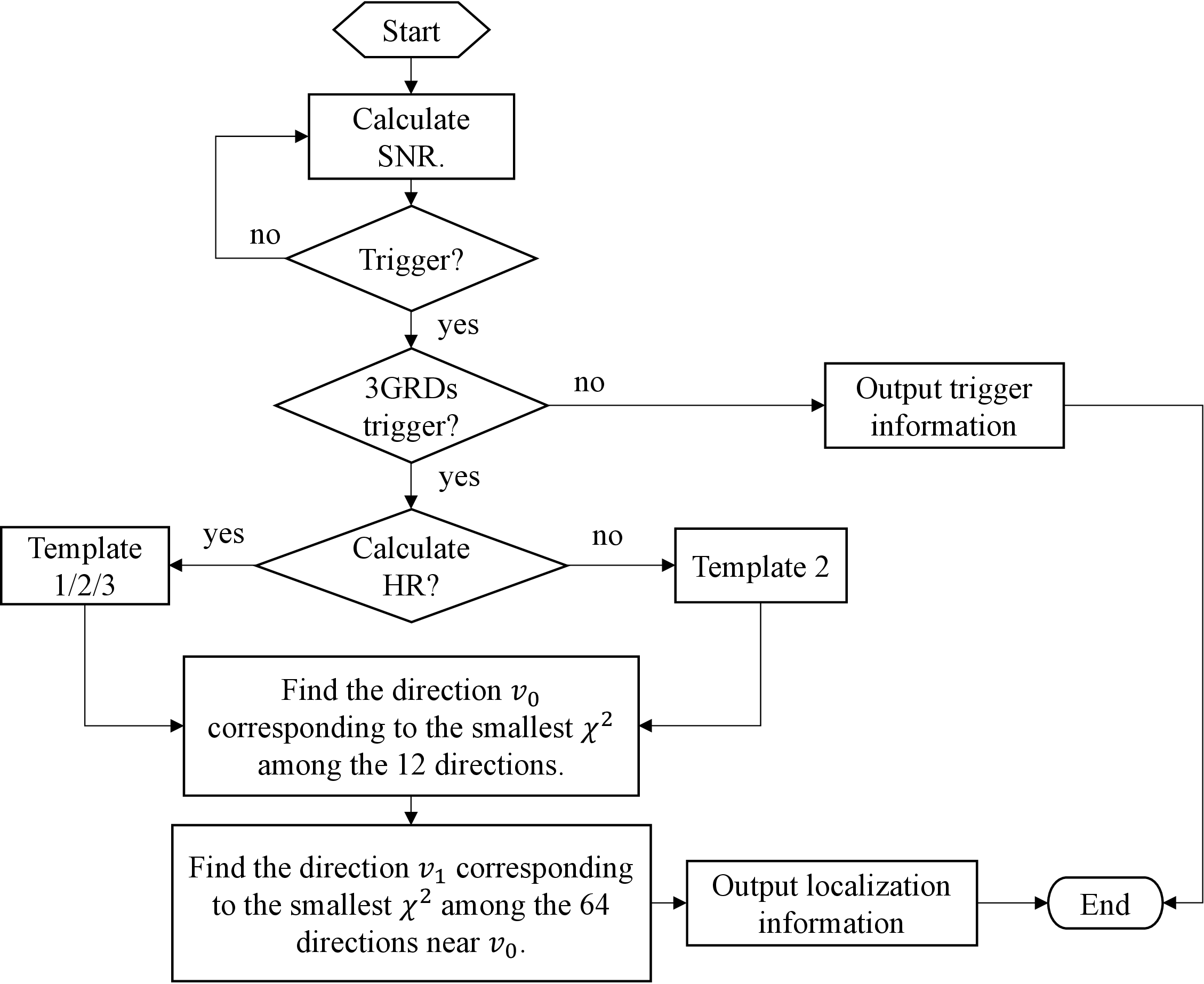}
    \caption{GRM in-orbit triggering and localization workflow.}
    \label{fig:local_process}
\end{figure}
The localization procedure post-triggering is summarized in Figure~\ref{fig:local_process}.
When only two GRDs exceed the SNR threshold, the system records trigger information (time, energy range, SNR) without localization information. localization are activated only when all three GRDs exceed the threshold. The software supports command-toggled hardness ratio calculations, defined as the count ratio between 15--50~keV and 50--300~keV. If disabled, precomputed reference templates 2 (see Table~\ref{tab:spectral_par}) are used to compute the $\chi^2$ statistic:

\begin{equation}\label{eq:chi2}
    \chi^{2} = \sum_{d=0}^{2} \sum_{c=0}^{1} \frac{S_{d,c} - f \cdot T_{d,c}}{S_{d,c} + B_{d,c}},
\end{equation}

\noindent where the normalization factor $f$ is derived from:

\begin{equation}\label{eq:norm}
    f = \frac{\sum_{d=0}^{2} \sum_{c=0}^{1} S_{d,c}}{\sum_{d=0}^{2} \sum_{c=0}^{1} T_{d,c}}.
\end{equation}

The detector index \(d\) (0: GRD01, 1: GRD02, 2: GRD03) and energy channel \(c\) (0: 15--50~keV, 1: 50--300~keV) jointly define the background counts \(B_{d,c}\), background-subtracted observed counts \(S_{d,c}\), and simulated template counts \(T_{d,c}\) derived from lookup tables.  

The template tables contain six columns, corresponding to the two energy channels across three GRDs. Each row represents a distinct incident direction, with 12 coarse (Healpix NSIDE=1, angular resolution $\sim60^\circ$) and 768 fine (Healpix NSIDE=8, resolution $\sim7^\circ$) directional bins. Templates were generated via Geant4 simulations using three Band function spectral parameter sets (Table~\ref{tab:spectral_par}), corresponding to distinct hardness.

\begin{table}
\begin{center}
\caption[]{Parameters of three typical hardness spectra.}\label{tab:spectral_par}
\begin{tabular}{ccccc}
\hline\noalign{\smallskip}
Template&Model &  $\alpha$      & $\beta$ & $E_\mathrm{peak}$(keV)                    \\
  \hline\noalign{\smallskip}
1&Soft  & -1.9 & -3.7    & 70 \\
2&Medium &-1.0 & -2.3   & 230  \\
3&Hard  &  0   &  -1.5  &  1000 \\
  \noalign{\smallskip}\hline
\end{tabular}
\end{center}
\end{table}
\subsection{Results of Triggering and Localization Testing}
\begin{figure}[!htbp]
    \centering
    \includegraphics[width=\linewidth]{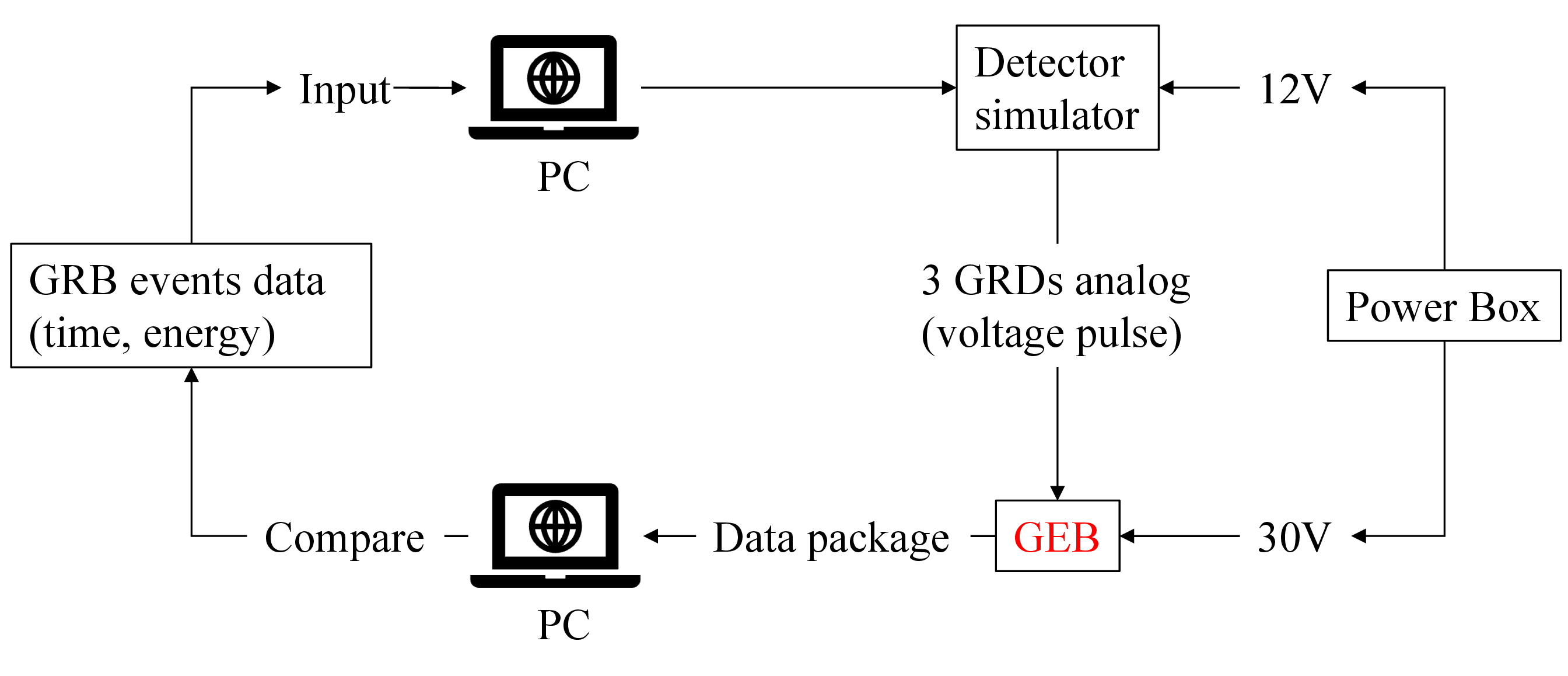}
    \caption{GRM ground test system schematic for validating in-orbit triggering and localization functions.}
    \label{fig:test_process}
\end{figure}

The GRM triggering and localization algorithms were validated through comprehensive ground-based software testing. As shown in Figure~\ref{fig:test_process}, the test framework simulates GRM's operational pipeline:  
\begin{enumerate}
    \item \textbf{Data Generation}: GRB signals and background noise were simulated using Geant4, with each simulated event containing precise timing and energy (15--5000~keV) information.
    \item \textbf{Signal Emulation}: The simulated events were converted into analog voltage pulses via a detector simulator~\citep{shi2022}.
    \item \textbf{Onboard Processing}: The GEB electronics digitized these pulses, executed real-time trigger logic, and localization calculations upon three GRD triggers.
\end{enumerate}

Based on different GRB templates and flux, We produced 52 GRB datasets using the Geant4. All 52 GRBs are capable of generating trigger packets, and the triggering moment occurs within 20 seconds after 100 seconds (with the GRB burst starting at 100 seconds), indicating that the triggering function is normal. Subsequently, we choose one GRB template to simulate data incidence from various directions (36 in total). The localization results are presented in the Table~\ref{tab:local_result}. It can be seen that the localization function is normal.

\begin{table}
\begin{center}
\caption[]{Localization results from different angles incident.}\label{tab:local_result}
\begin{tabular}{cccc}
\hline
  &\multicolumn{3}{c}{Zenith($^{\circ}$)} \\
Azimuth($^{\circ}$) & $15$ & $30$ & $45$ \\
  \hline
0    &(17.6, 345.0)  & (29.6, 351.0) & (48.2, 5.6)\\
60   &(17.6, 75.0)  & (35.7, 67.5)   & (48.2, 61.6) \\
120  &(11.7, 112.5) & (23.6, 123.7)  & (41.9, 109.3) \\
180  &(17.6, 165.0) & (29.6, 171.0)  & (48.2, 196.9)\\
240  &(11.7, 247.5) & (29.6, 243.0)  & (48.2, 241.9) \\
300  &(17.6, 285.0) & (29.6, 279.0)  & (48.2, 309.4)\\
\hline
&60&75&90\\
\hline
0&--&(80.4, 10.3)&(80.4, 28.1)\\
60&--&--&--\\
120&(60.0, 129.4)&--&--\\
180&--&--&--\\
240&(70.5, 253.1)&--&--\\
300&--&--&--\\
\hline
\end{tabular}
\end{center}
\tablecomments{0.5\textwidth}{The '--' signifies that only two GRDs have been triggered, no localization results.}
\end{table}
\section{Ground-based GRB Localization Algorithm}
\label{sec:xband}
\subsection{Methodology}
The on-orbit localization method used on GRM satellites is similar to BATSE's LOCBURST~\citep{pendleton1999} (The BATSE Burst Location Algorithm) and GBM's DOL~\citep{connaughton2015} (Daughter Of Location), which involves using the Band spectral function to simulate the counts from GRBs in different directions on the three GRD detectors and store them in the memory. After the triggering, the chi-square value is computed, and the angle corresponding to the minimum chi-square value is obtained as the in-orbit localization result. This method is resource-efficient and fast in computation. However, GRM performs onboard localization only when all three GRD detectors are triggered, and the localization relies on the count rates from the trigger timescale (0.1, 1, or 4 seconds), resulting in low photon statistics. Consequently, the onboard localization efficiency is limited and the positional errors are large. Furthermore, the use of fixed spectral parameters in the template-matching approach introduces additional systematic errors \citep{berlato2019}. To mitigate such biases, the BALROG (BAyesian Location Reconstruction Of GRBs) framework was developed for ground-based localization, which has been shown to effectively reduce systematic uncertainties \citep{burgess2018, zhao2023, zhao2024, liao2020}.

The GRM ground-based localization algorithm employs a likelihood-based method similar to BALROG. The first step is to construct the likelihood function. Within a time interval $t$, the number of photons detected by detector $d$ in energy channel $c$ follows a Poisson distribution. Therefore, the probability of detector $d$ in the direction $\Omega$ and spectral parameter $\Psi$ can be expressed as:
\begin{equation}\label{eq:prob_pd}
    p_{d}(\Psi,\Omega)=\prod_{c=1}^{N_{c}}{\frac{(T_{d,c}t)^{\lambda_{d,c}}e^{-T_{d,c}t}}{\lambda_{d.c}!}},
\end{equation}
When the count is high enough, it can be simplified to Gaussian distribution, as shown below.
\begin{equation}\label{eq:prob_pd}
    p_{d}(\Psi,\Omega)=\prod_{c=1}^{N_{c}}{\frac{1}{\sqrt{2\pi}\sigma_{d,c}}exp(-\frac{(S_{d,c}-T_{d,c})^{2}}{2\sigma_{d,c}^{2}})},
\end{equation}
Therefore, the likelihood function $L(\Psi,\Omega)$ can be expressed as equation~\ref{eq:likehood}:
\begin{equation}\label{eq:likehood}
    L(\Psi,\Omega)=\prod_{d=1}^{N_{d}}p_{d}(\Psi,\Omega).
\end{equation}

By seeking the parameters corresponding to the maximum value of the likelihood function, the optimal direction for localization can be obtained. When there are few parameters, the optimal value can be found by traversing the entire parameter range, but when there are too many parameters, the time and computational resources required for traversal become significant. Here, we employ the method of Markov chain to sample the parameter space and obtain the confidence interval of the parameters. The likelihood function reflects the probability of detecting $\lambda$ counts when the parameters are ($\Psi$,$\Omega$), denoted as P($\lambda\Psi$,$\Omega$). According to the Bayes' theorem, we can infer the probability distribution of the parameter space P($\Psi$,$\Omega|\lambda$) when detecting the count rate $\lambda$
\begin{equation}
    P(\Psi,\Omega|\lambda)=\frac{P(\lambda|\Psi,\Omega)P(\Psi,\Omega)}{\sum (P(\lambda|\Psi,\Omega)P(\Psi,\Omega)}.
\end{equation}

Assuming the prior distribution $P(\Psi,\Omega)$ of the parameter space is a uniform distribution, the denominator integral yields a fixed value, so $P(\Psi,\Omega|\lambda) \propto L(\Psi,\Omega)$. Once the form of the probability density function is known, the Monte Carlo Markov Chain (MCMC) method can be employed for sampling. Commonly used MCMC methods include the Metropolis-Hastings (M-H) sampling and Gibbs sampling, as well as some improved methods. In this work, we performed MCMC sampling using the emcee package~\citep{emcee2013}. To assess convergence, we first discarded the initial 20$\%$ of samples as burn-in. We then computed the autocorrelation time ($\tau$) for the remaining chain and verified that the chain length exceeded $50\times\tau$. In addition, the chains were visually inspected to ensure stability. Samples satisfying these criteria were considered effectively independent and used for posterior inference and parameter estimation.
\subsection{Validation with Simulated GRBs}\label{sec:4-2}

To further investigate the impact of spectral assumptions on localization accuracy, we simulated two kinds of GRBs with different fluences of $1\times$10$^{-5}$ erg/cm$^{2}$ and $1\times$10$^{-6}$ erg/cm$^{2}$ (integrated over 1~s), respectively. The simulated burst adopts a Comptonized (CPL) spectrum with $\alpha$=1 and E$_\mathrm{peak}$=200 keV, and is incident from a direction of ($\theta$=30$^{\circ}$, $\phi$=120$^{\circ}$) in GRM coordinates. Using three different approaches—fixed: fixing the spectral parameters to their true values; free: allowing all spectral parameters to vary freely; and offset: fixing the spectral parameters to incorrect values—we sampled the directional parameters.

\begin{figure}[!htbp]
    \centering
    \includegraphics[width=\linewidth]{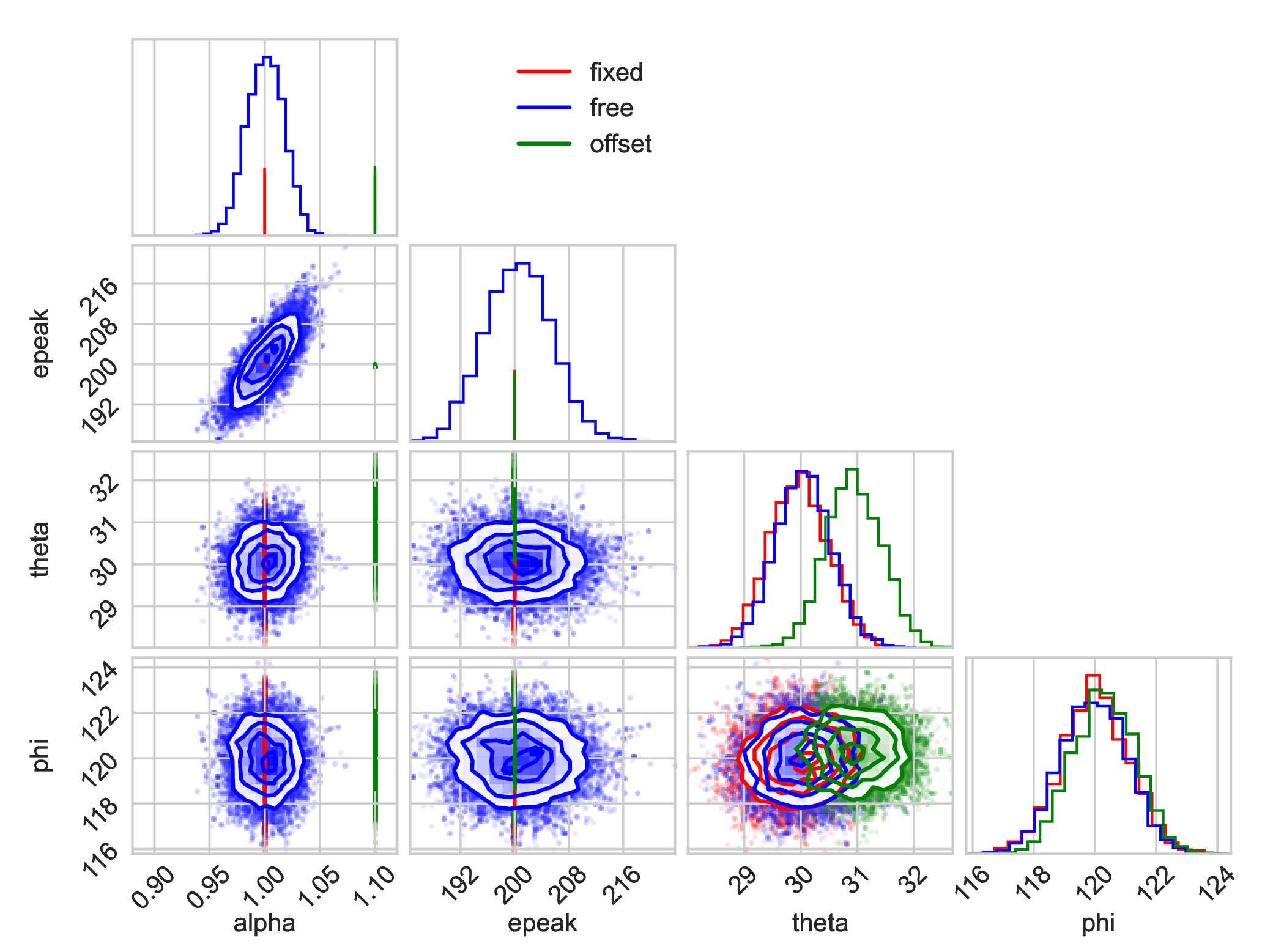}
    \includegraphics[width=\linewidth]{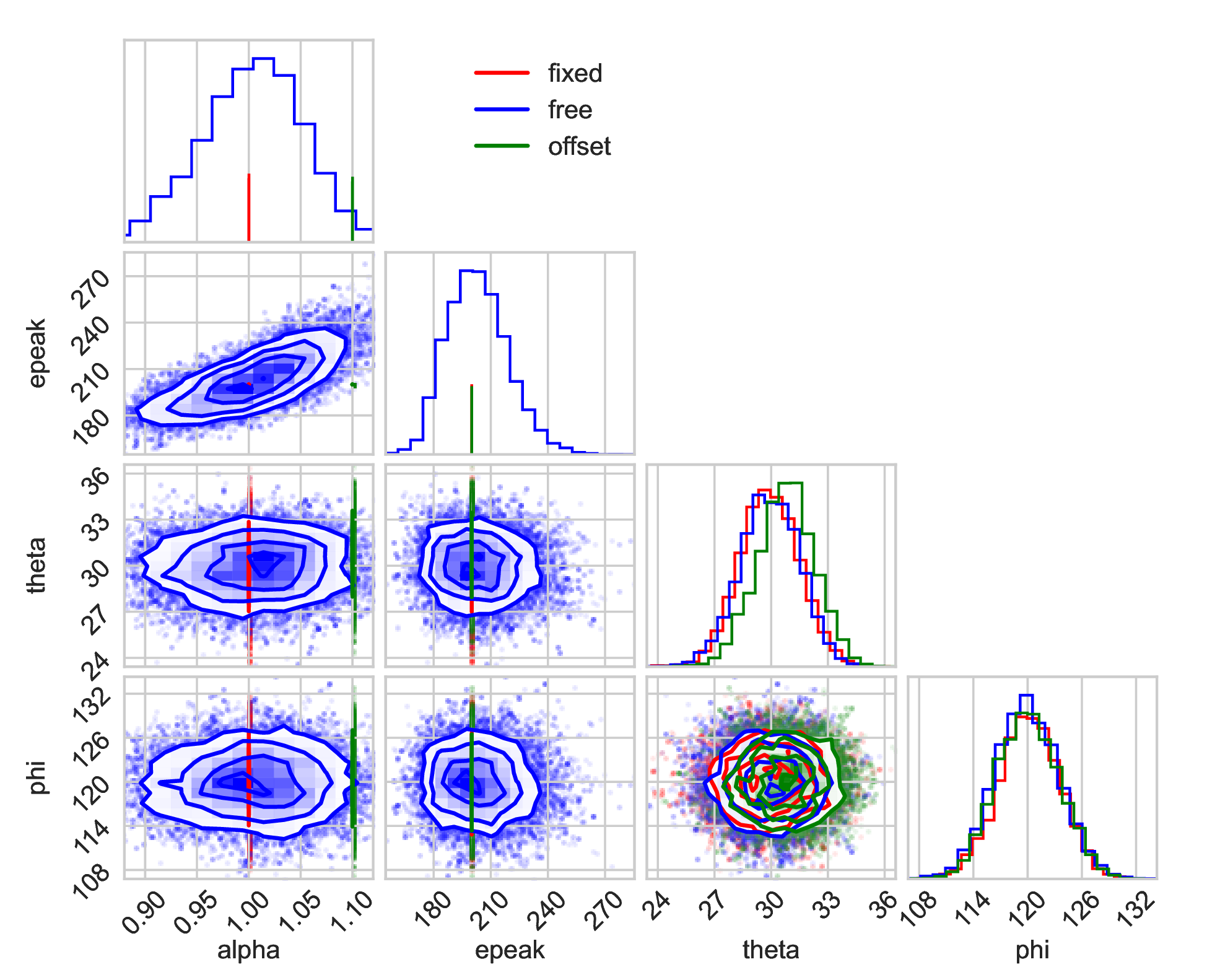}
    \caption{Impact of spectral assumptions on localization for simulated GRBs with fluences of $1\times$10$^{-5}$ erg/cm$^{2}$ (top) and $1\times$10$^{-6}$ erg/cm$^{2}$ (bottom). Systematic errors are evident for bright GRBs where statistical uncertainties are small, but become negligible for faint GRBs where statistical errors dominate.}
    \label{fig:fix_free_offset_compare}
\end{figure}

The results, shown in Figure~\ref{fig:fix_free_offset_compare}, demonstrate that incorrect spectral assumptions introduce significant systematic errors in localization. The systematic bias is most pronounced when statistical uncertainties are small (i.e., for bright bursts), whereas for faint GRBs where statistical errors dominate, the bias becomes negligible.

The systematic errors introduced by misspecified spectral parameters, while significant for bright GRBs, have a reduced impact on faint bursts. To account for these effects in our analysis, the GRM ground-based localization pipeline adopts a joint sampling approach that simultaneously fits both the spectral and localization parameters. Unless otherwise stated, all ground-based localizations presented in the following sections are obtained using this method.

\section{Localization Analysis}
\subsection{Localization Performance with Simulations}
\begin{figure}[!htbp]
    \centering
    \includegraphics[width=\linewidth]{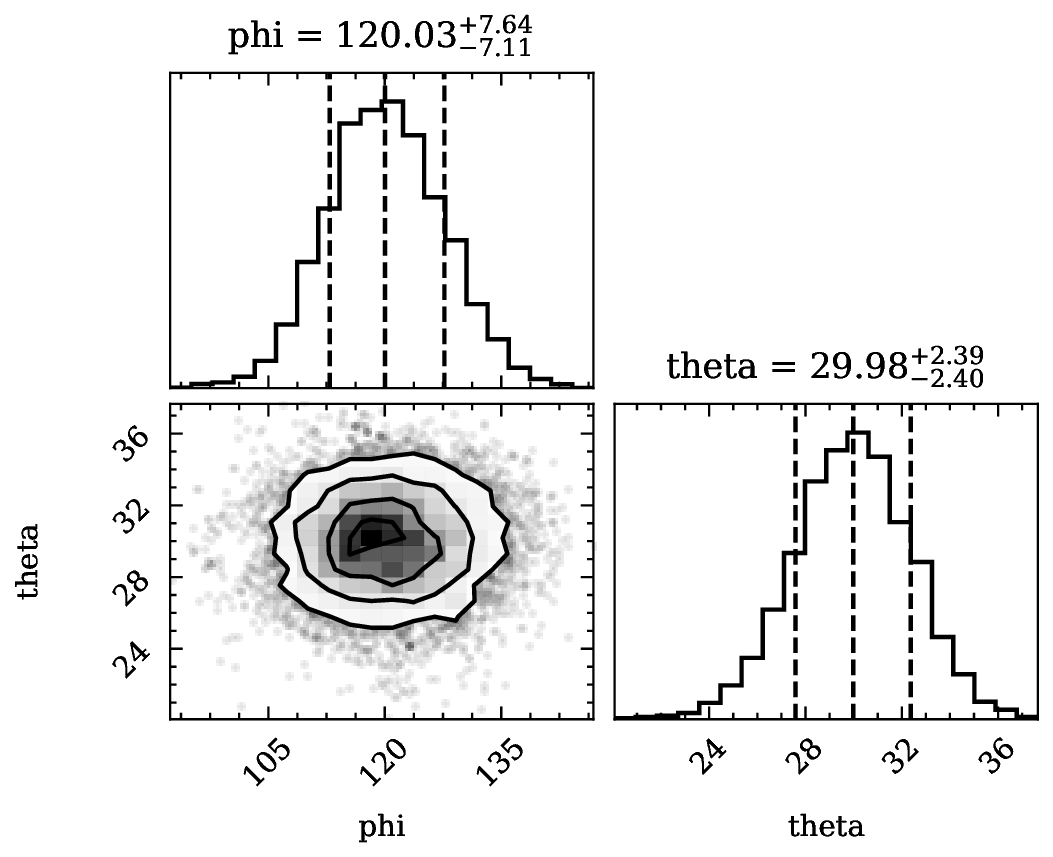}
    \caption{The GRB localization results in the (30$^{\circ}$,120$^{\circ}$) direction.}
    \label{fig:local_result_30_120}
\end{figure}

\begin{figure}[!htbp]
    \centering
    \includegraphics[width=\linewidth]{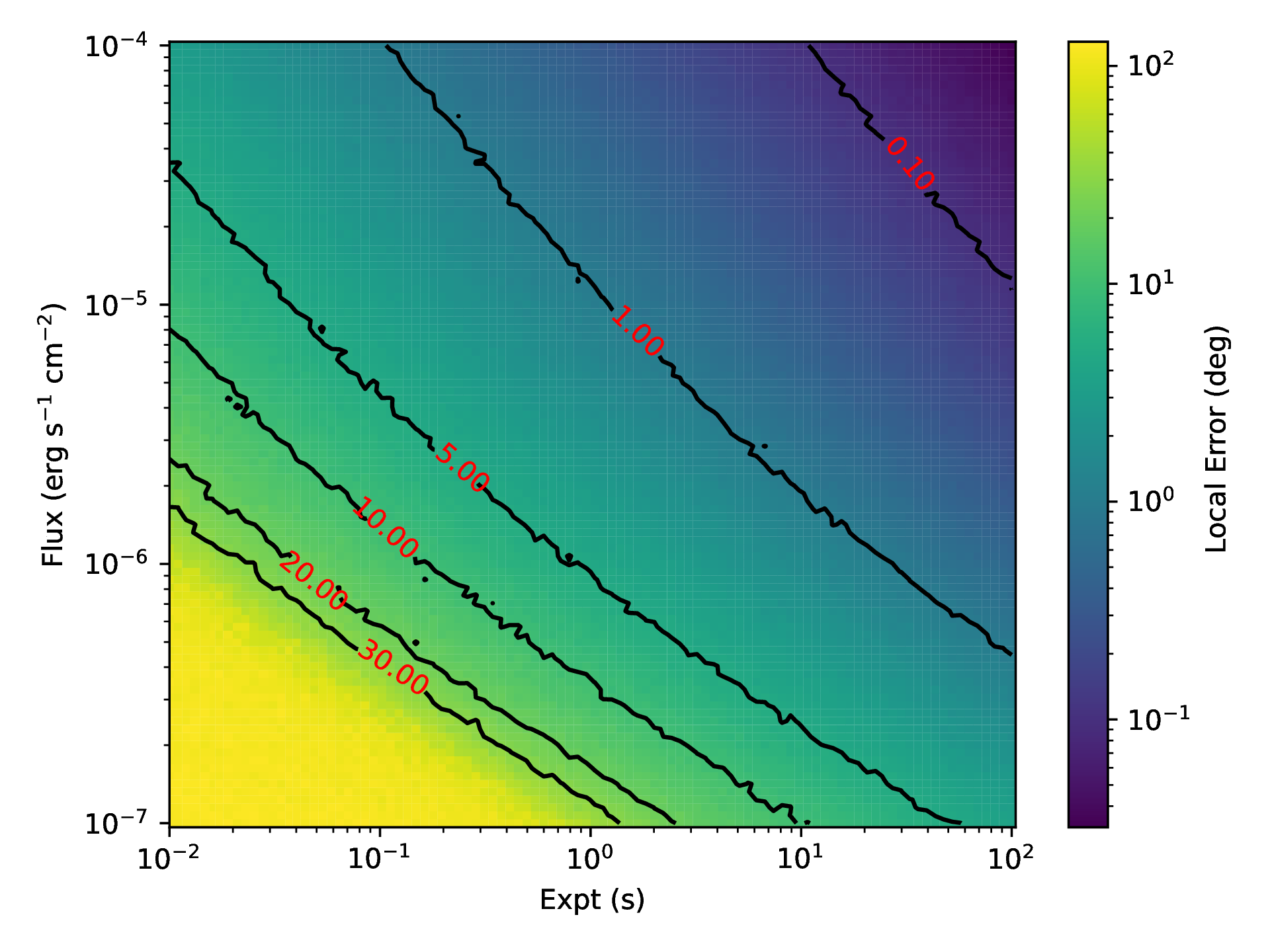}
    \caption{Localization error estimates for different fluxes as a function of burst duration. The black lines from top to bottom correspond to 0.1$^\circ$, 1$^\circ$, 5$^\circ$, 10$^\circ$, 20$^\circ$ and 30$^\circ$ respectively.}
    \label{fig:local_err_expo_flux}
\end{figure}

Using the method in sec~\ref{sec:xband}, GRB incident in the direction of (30°, 120°)  were simulated. The location results indicate that for a GRB with a fluence of 1×10$^{-6}$ erg cm$^{-2}$ s$^{-1}$, spectral parameters $\alpha$=-1.0, $\beta$=-2.3, E$_{peak}$=230~keV, and a duration of 1 second, the 1$\sigma$ error interval is $\theta$=$29.98_{-2.40}^{+2.39}$ $\phi$=120.03$_{-7.11}^{+7.64}$, meeting the requirement of less than 5$^\circ$(See Fig~\ref{fig:local_result_30_120}). To further evaluate the statistical localization errors for GRBs with different fluences and durations, we present in Fig.~\ref{fig:local_err_expo_flux} the localization uncertainties as functions of both flux and burst duration, providing a comprehensive assessment of the expected performance across a wide range of burst flux and timescales.

\begin{figure*}[!htbp]
    \centering
    \includegraphics[width=\linewidth]{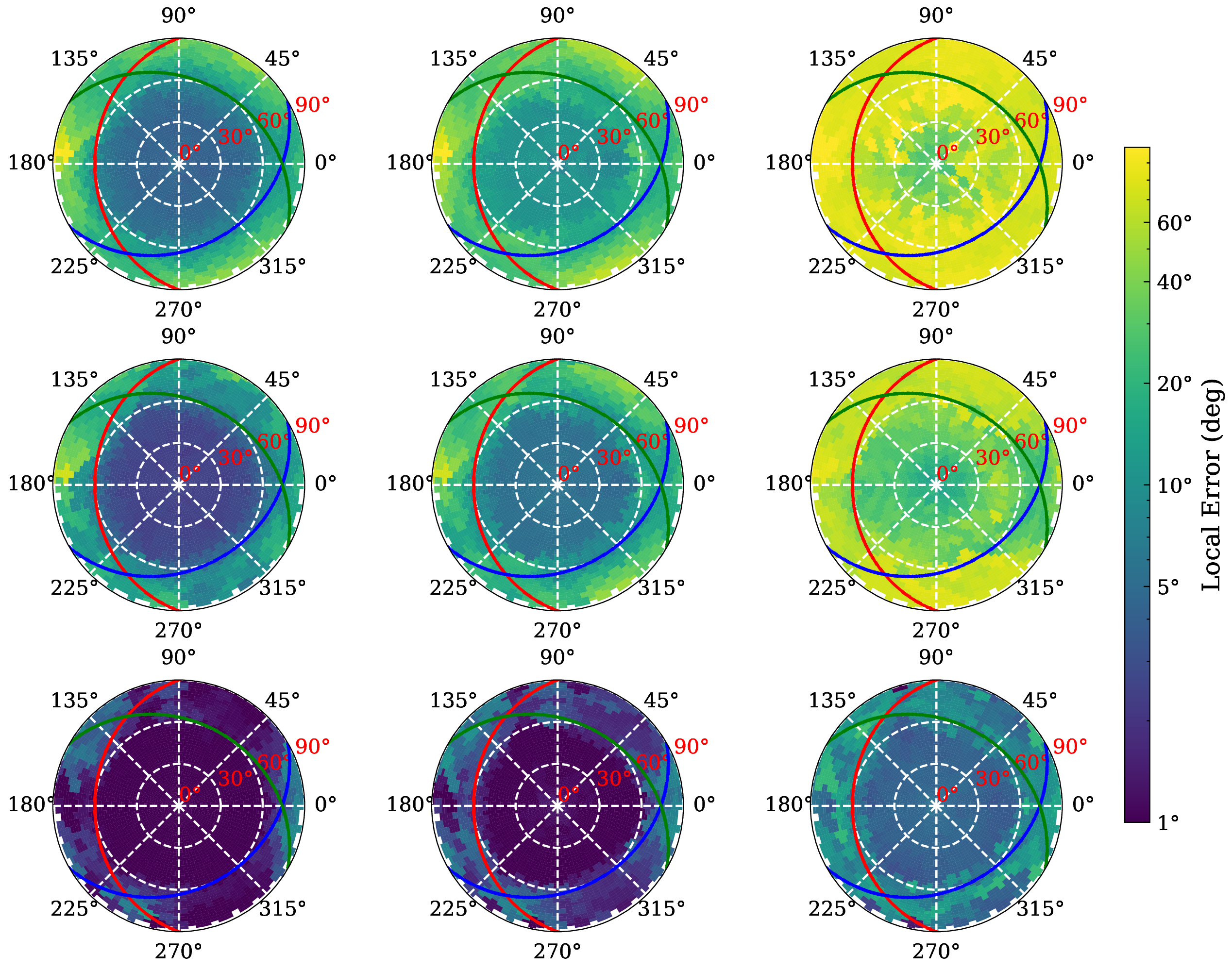}
    \caption{Localization statistical errors for GRBs with different spectral hardness and incident directions. From left to right, the three columns correspond to soft, medium, and hard spectra (see Table~\ref{tab:spectral_par}), respectively. From top to bottom, the three rows represent fluences of 5$\times$10$^{-7}$, 1$\times$10$^{-6}$,1$\times$10$^{-5}$ erg/cm$^{2}$ in 1~s. The red, green, and blue lines indicate the FoV boundaries of the three GRDs.}
    \label{fig:local_err}
\end{figure*}

We performed simulations and localization for a grid of 368 incident directions across three different fluence levels (5$\times$10$^{-7}$, 1$\times$10$^{-6}$,1$\times$10$^{-5}$ erg/cm$^{2}$ in 1~s). The resulting statistical localization errors are shown in Figure~\ref{fig:local_err}. for medium-spectrum GRBs with a typical fluence of 1$\times$10$^{-6}$erg/cm$^{2}$, the 1$\sigma$ localization errors within the common field of view of the three detectors ($\theta<$60$^{\circ}$) are generally below 5$^{\circ}$.

\subsection{Localization Results with GRB Data}
SVOM was successfully launched on June 22, 2024, and the GRM started operating on June 27. Fortunately, the GRM detected the first three GRBs on June 27 (GRB 240627B), June 29 (GRB 240629A), and July 2 (GRB 240702A), respectively~\citep{GCN36805}. GRB 240629A is used to validate the localization algorithm for the GRB is relatively bright.

\begin{figure}[!htbp]
    \centering
    \includegraphics[width=\linewidth]{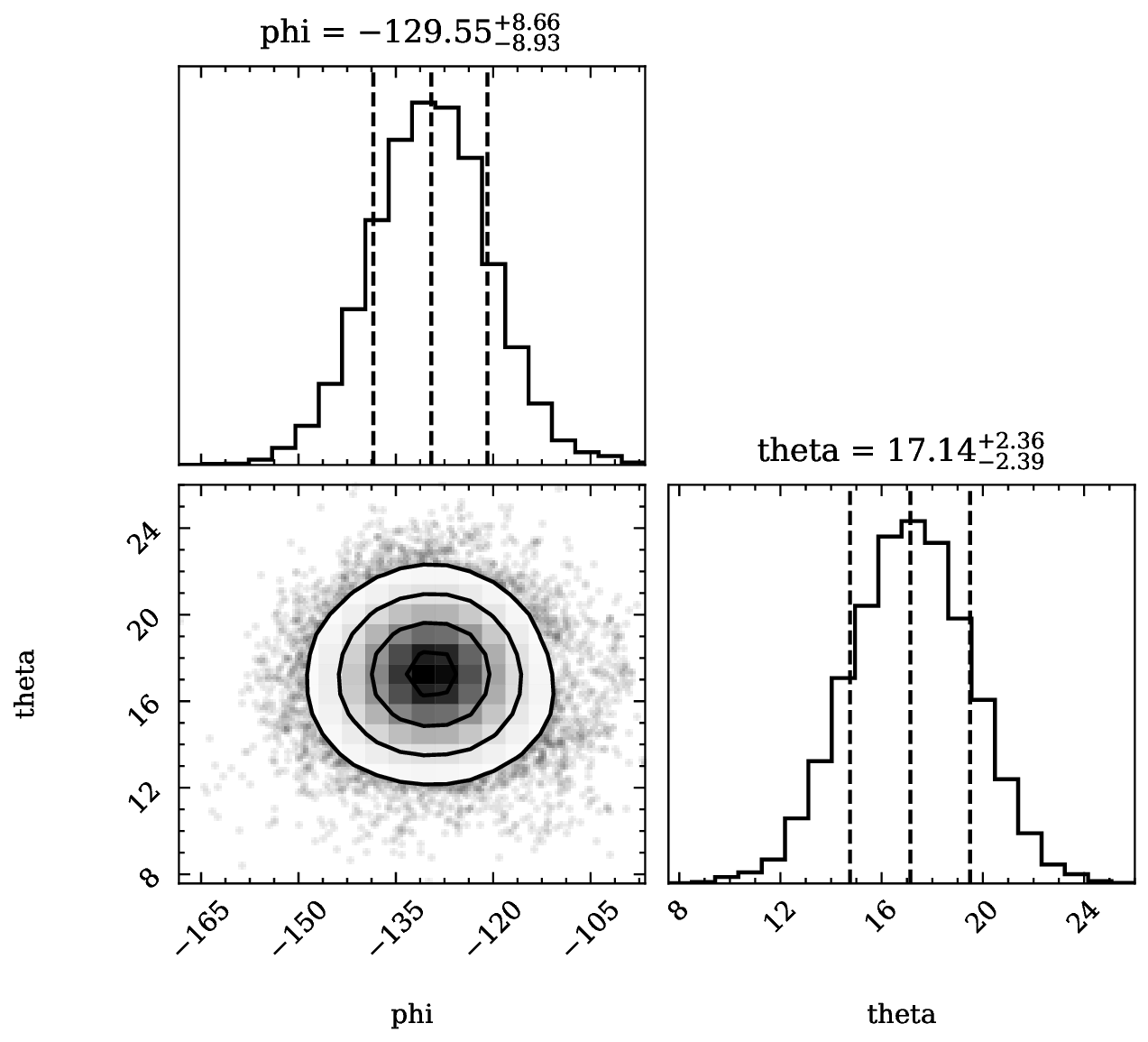}
    \caption{The distribution of theta and phi of GRB240629A in the GRM's coordinate system.}
    \label{fig:local_grb240629A}
\end{figure}
\begin{figure}[!htbp]
    \centering
    \includegraphics[width=0.9\linewidth]{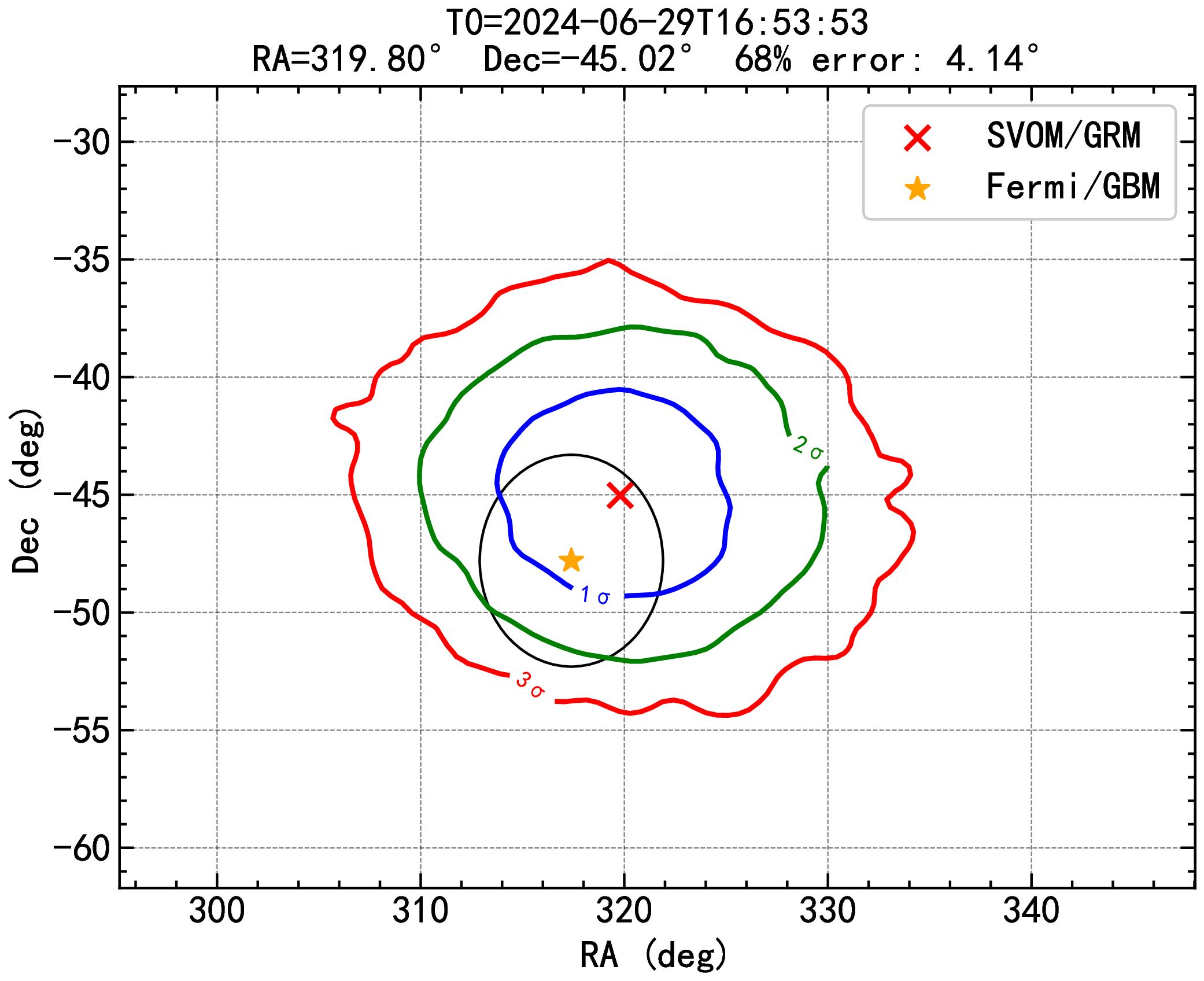}
    \caption{The GRM's localization results for GRB 240629A. The yellow star and black circle represent the GBM localization result and its associated error, respectively.}
    \label{fig:local_grb240629A_sky}
\end{figure}

According to the satellite's attitude quaternion, we transform theta and phi to the J2000 coordinate system. As shown in the Fig~\ref{fig:local_grb240629A_sky}, the three circles represent the 1, 2, and 3 sigma confidence region, with the maximum probability density position at (319.80$^{\circ}$, -45.02$^{\circ}$), which is consistent with the localization result of Fermi/GBM (317.4$^{\circ}$, -47.8$^{\circ}$)~\cite{GCN36787}.

\begin{figure*}[htbp]
  \centering
  \begin{subfigure}{0.31\textwidth}
  \includegraphics[width=\linewidth]{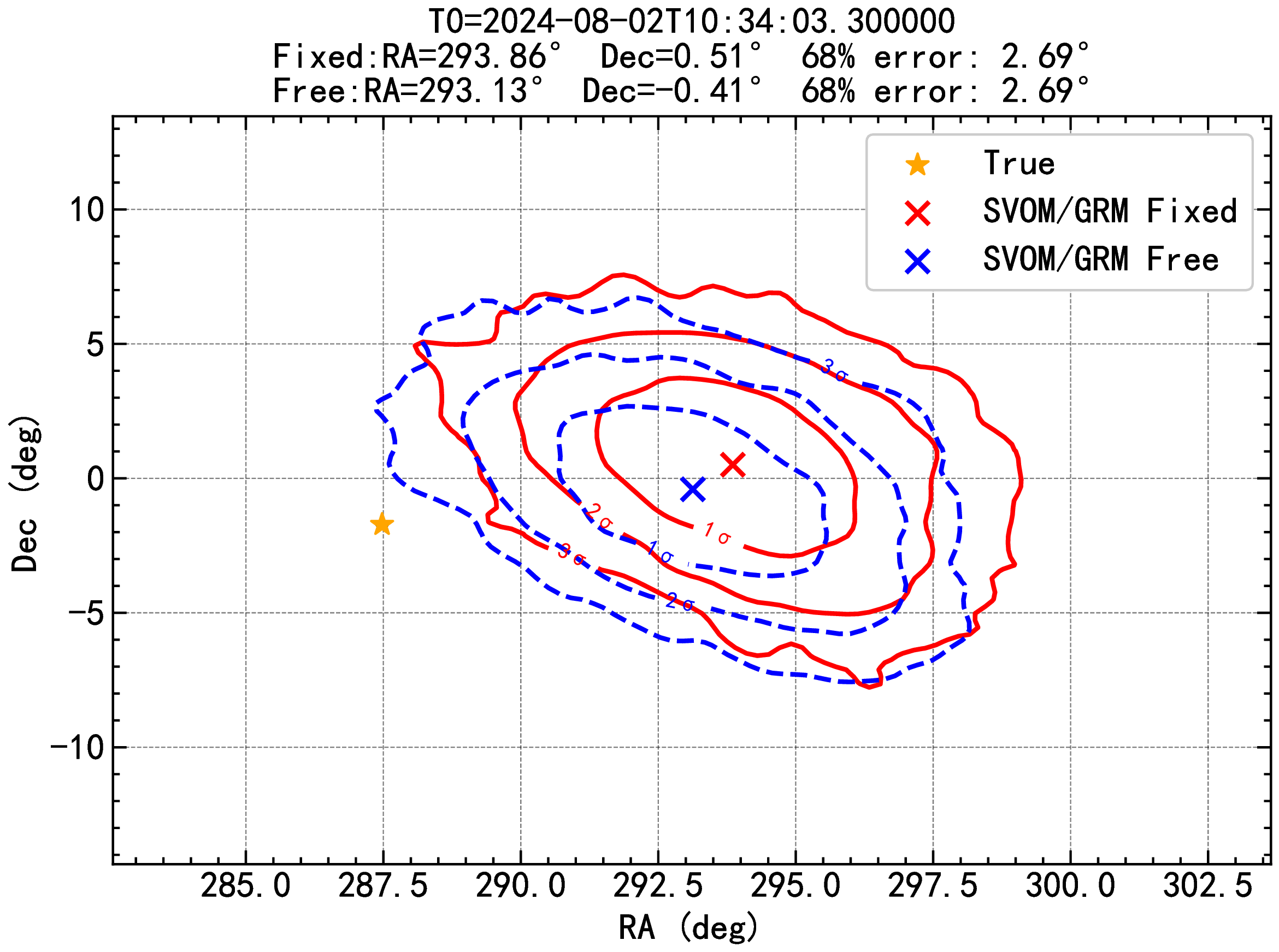}
  \caption{GRB 240802A}
  \end{subfigure}
\begin{subfigure}{0.31\textwidth}
  \includegraphics[width=\linewidth]{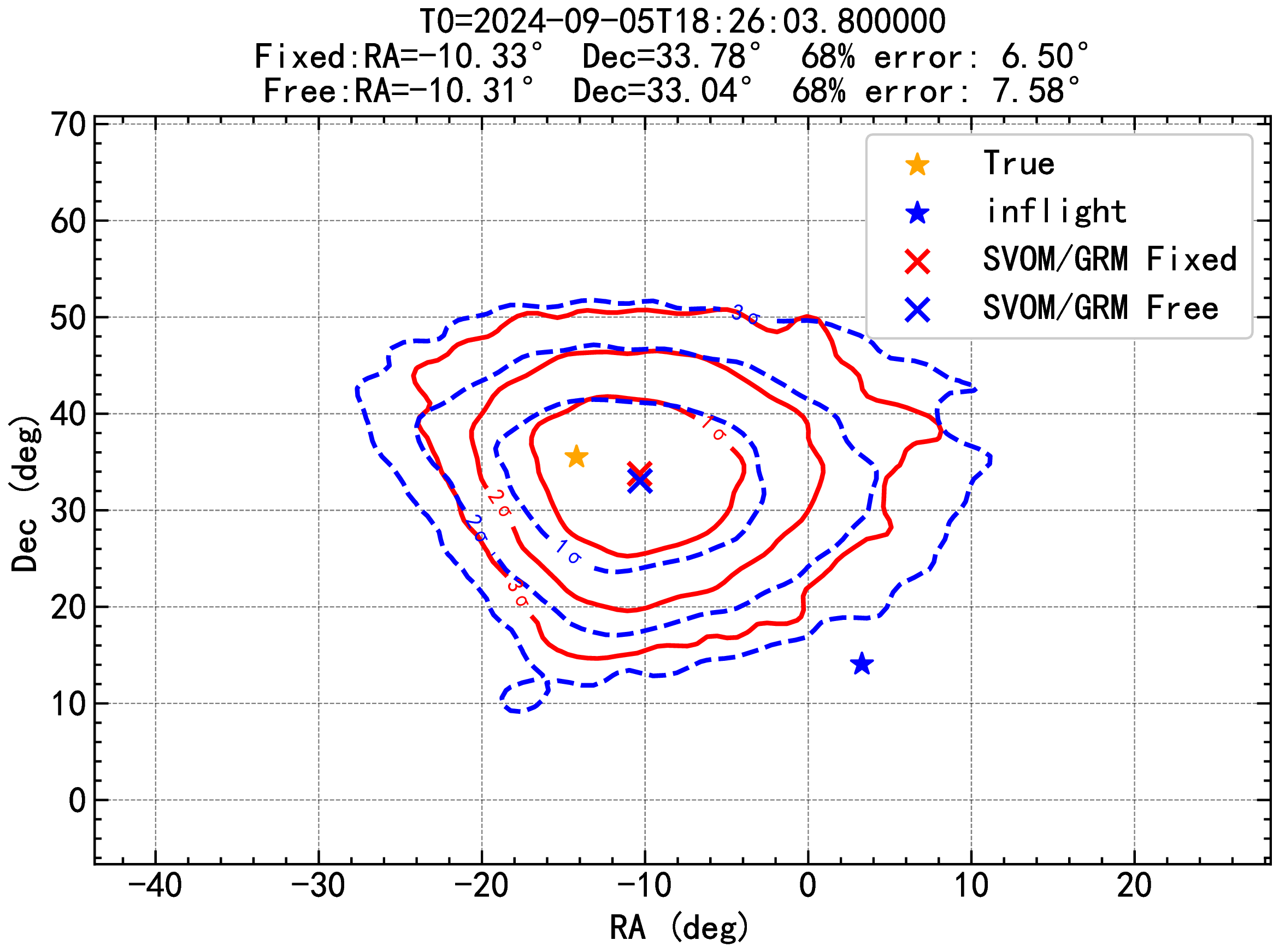}
  \caption{GRB 240905E}
  \end{subfigure}
   \begin{subfigure}{0.31\textwidth}
  \includegraphics[width=\linewidth]{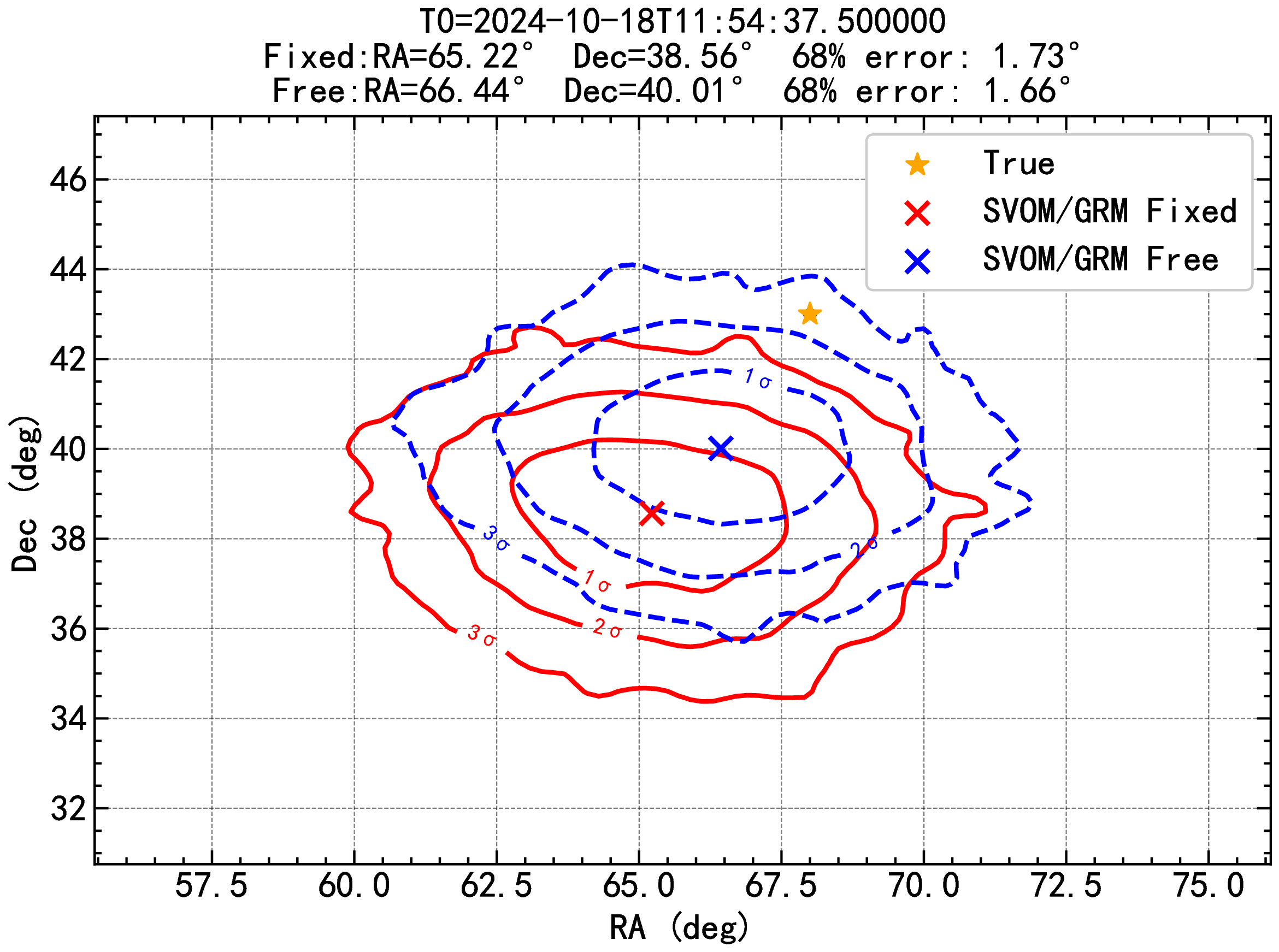}
  \caption{GRB 241018A}
  \end{subfigure}
  \\
   \begin{subfigure}{0.31\textwidth}
  \includegraphics[width=\linewidth]{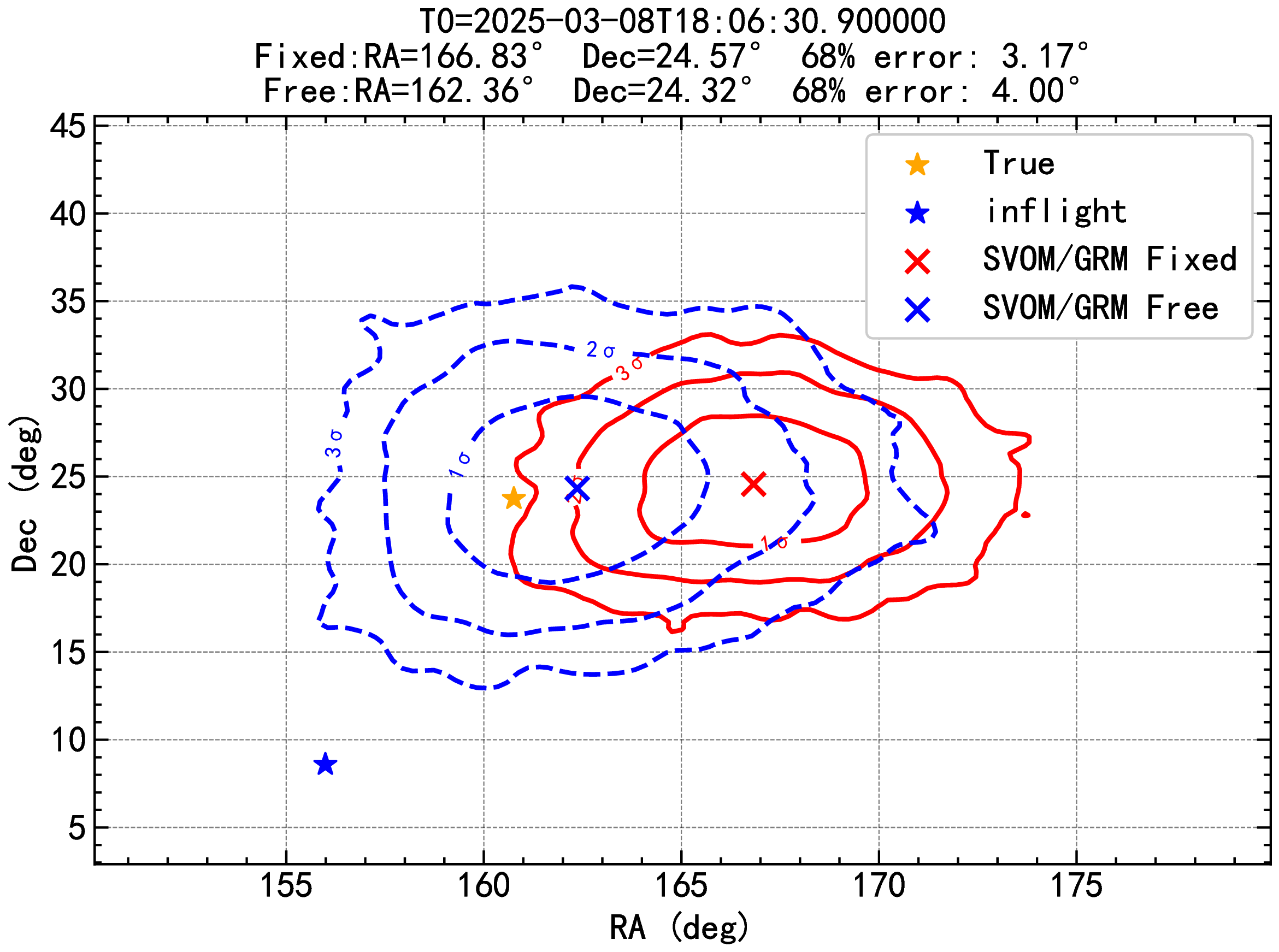}
  \caption{GRB 250308A}
  \end{subfigure}
   \begin{subfigure}{0.31\textwidth}
 \includegraphics[width=\linewidth]{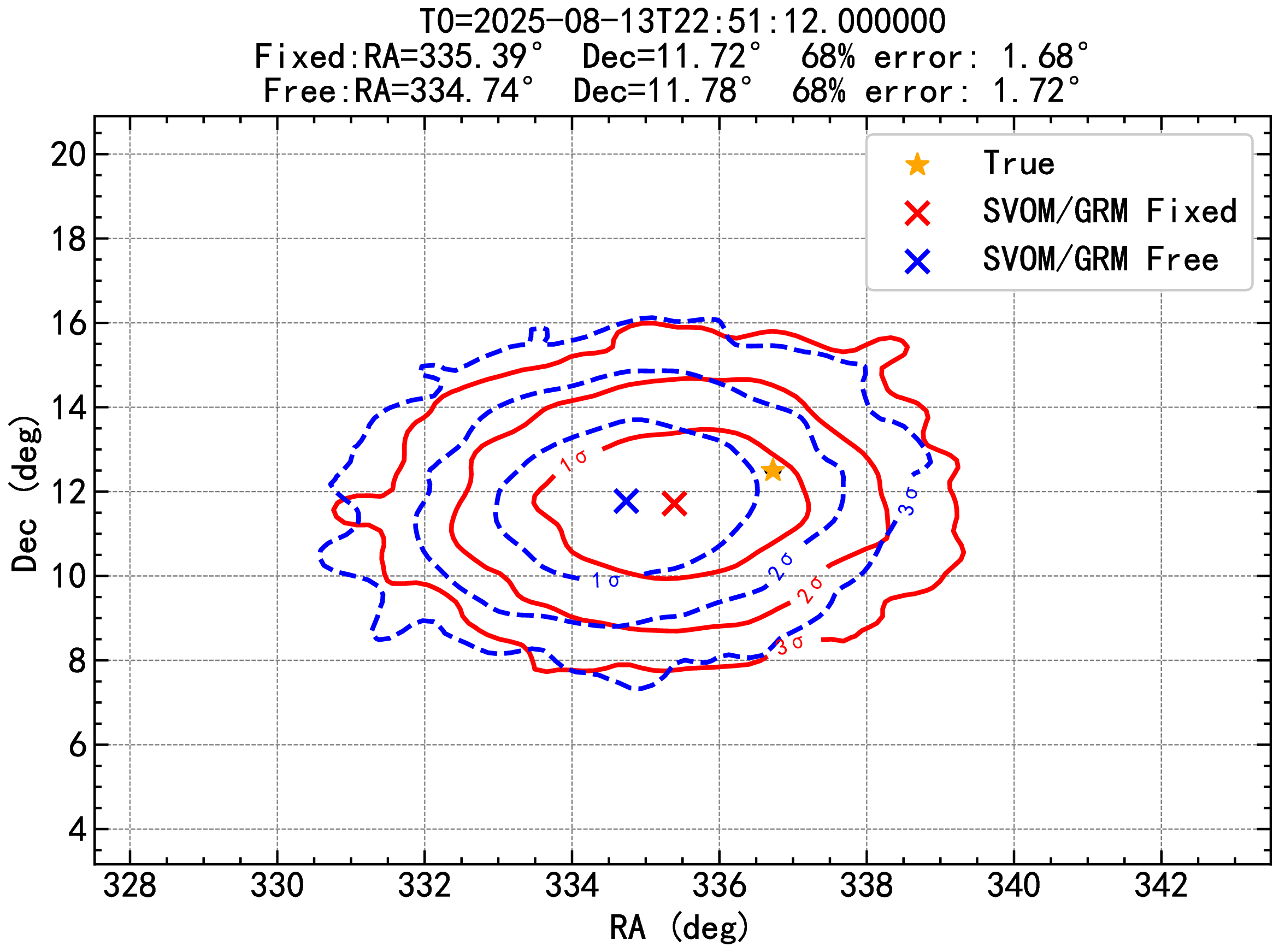}
  \caption{GRB 250813B}
  \end{subfigure}
  \begin{subfigure}{0.31\textwidth}
 \includegraphics[width=\linewidth]{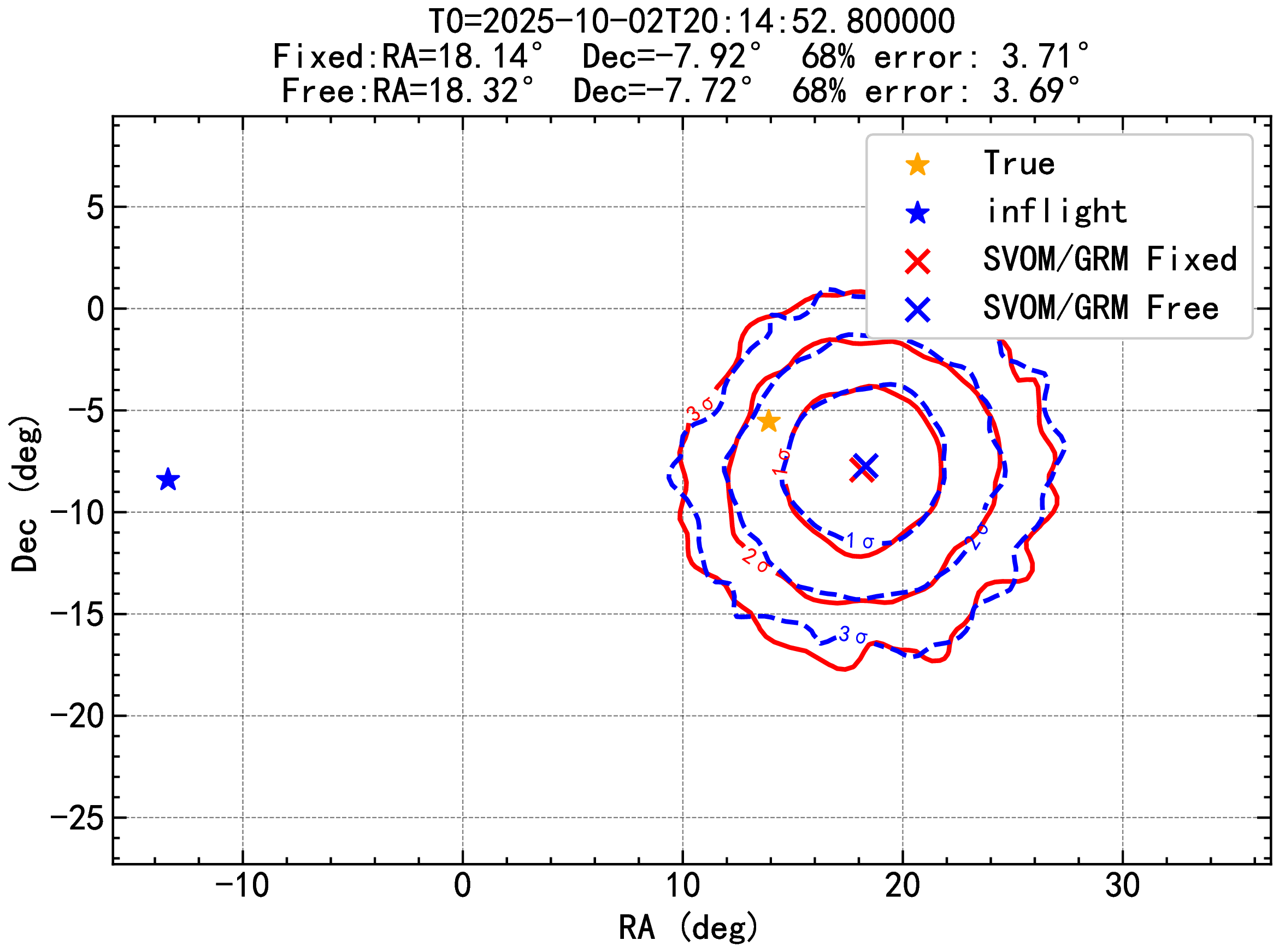}
 \caption{GRB 251002A}
  \end{subfigure}
  \caption{Localization results for six GRBs using different methods. The yellow stars indicate the true (reference) positions of the GRBs. Red crosses with solid lines represent the ground-based localization results obtained with fixed spectral parameters, while blue crosses with dashed lines correspond to the ground-based joint fitting results where both spectral and localization parameters are sampled. For GRB 240905E, GRB 250308A, and GRB 251002A, the onboard real-time localization results are also shown.}
  \label{fig:local_comparison}
\end{figure*}

By December 2025, GRM had detected over 100 GRBs since its commissioning. However, only a limited number of bright bursts have precise independent localizations. Table~\ref{tab:grb_localization_results} presents six GRBs with both onboard and ground-based localization results. As shown in Figure~\ref{fig:local_comparison}, the ground-based pipeline significantly improves upon the onboard positions.

Simulations in Section~\ref{sec:4-2} demonstrate that fixed spectral parameters can introduce systematic errors. Nevertheless, for real GRBs, the joint fitting approach does not consistently outperform the fixed-spectrum method. In particular, the localization of GRB~240802A deviates from its reference position by more than 3$\sigma$, exceeding the expected confidence region. This indicates that additional systematic effects—such as detector response uncertainties, atmospheric scattering, among others—may also contribute to the residual errors. A larger sample of well-localized GRBs is required to statistically characterize these biases. Future work will focus on deriving an overall systematic error estimate for GRM.
\begin{table*}[]
\centering
\caption{Localization results for GRBs detected by GRM.}
\label{tab:grb_localization_results}
\begin{tabular}{lcccccc}
\hline
\multirow{2}{*}{GRB name}&\multirow{2}{*}{Time(UTC)}&\multirow{2}{*}{position}&\multicolumn{3}{c}{local results}\\
\cline{4-6}
&&&inflight&fixed&free\\
\hline
GRB 240802A&2024-08-02T10:34:03.3&(287.5, -1.7)&--&(293.9, 0.5)&(293.1, -0.4)\\
GRB 240905E&2024-09-05T18:26:03.8&(345.8, 35.5)&(3.3, 14.1)&(-10.4, 33.7)&(-10.5, 33.2)\\
GRB 241018A&2024-10-18T11:54:37.5&(68.0, 43.0)&--&(65.2, 38.5)&(66.4, 40.0)\\
GRB 250308A&2025-03-08T18:06:30.9&(160.8, 23.7)&(156.0, 8.6)&(166.9, 24.5)&(162.4, 24.3)\\
GRB 250813B&2025-08-13T22:51:12.0&(336.7, 12.5)&--&(335.4, 11.7)&(334.7, 11.8)\\
GRB 251002A&2025-10-02T20:14:52.8&(13.9, -5.5)&(346.6, -8.4)&(18.2, -7.9)&(18.3, -7.7)\\
\hline
\end{tabular}
\tablecomments{\textwidth}{The '--' indicates that only two GRDs were triggered, so no onboard localization result is available.The positions are from Swift/BAT, IPN, or SVOM/ECLAIRs and are assumed to be the true GRB locations.}
\end{table*}

\section{Summary}\label{sect:discussion}
 In this article, we provided a detailed introduction to the on-orbit triggering and localization algorithm of GRM. We tested the triggering and localization functionality using the simulator, and the test results indicated that the triggering and localization functions of GRM can operate normally. Additionally, addressing the shortcomings of the on-orbit localization algorithm, we designed a ground-based localization algorithm by constructing likelihood functions with energy spectra and directions as variables, and used the MCMC method to compute the posterior distribution of the parameters. Using this method, we simulated the localization statistical errors of GRBs with different hardness spectrum and found that softer spectra has higher photon counts which lead to smaller localization statistical errors. The localization results of soft spectra meet the requirement of less than 5$^{\circ}$. Finally, we used this algorithm to localize the first relatively bright GRB detected by GRM, and the localization results demonstrate that this algorithm can local the GRB. Given the challenges in localizing faint or off-axis GRBs, further refinement of the current algorithm is necessary. Future enhancements may incorporate timing-based triangulation using inter-spacecraft delays, as well as joint spectral and positional fits with external instruments such as Fermi/GBM, Swift/BAT, and GECAM. With continued development of the GRM ground pipeline and future multi-satellite collaborative efforts, GRM is expected to play a meaningful role in enabling rapid multi-messenger follow-up observations.
\begin{acknowledgements}
The Space-based multi-band Variable Objects Monitor (SVOM) is a joint Chinese-French mission led by the Chinese National Space Administration (CNSA), the French Space Agency (CNES), and the Chinese Academy of Sciences (CAS). We gratefully acknowledge the unwavering support of NSSC, IAMCAS, XIOPM, NAOC, IHEP, CNES, CEA, and CNRS. The authors are grateful for support from the National Key R$\&$D Program of China (grant Nos. 2024YFA1611700, 2024YFA1611701, 2024YFA1611703), the National Natural Science Foundation of China (Grant No. 12494572, 12494570, 12273042, 11961141013 and 12333007), the Strategic Priority Research Program of the Chinese Academy of Sciences (Grant No. XDB0550300), and China's Space Origins Exploration Program.
\end{acknowledgements}
\label{lastpage}
\bibliographystyle{raa}
\bibliography{bibtex}

@ARTICLE{klebesadel1973,
       author = {{Klebesadel}, Ray W. and {Strong}, Ian B. and {Olson}, Roy A.},
        title = "{Observations of Gamma-Ray Bursts of Cosmic Origin}",
      journal = {\apjl},
         year = 1973,
        month = jun,
       volume = {182},
        pages = {L85},
          doi = {10.1086/181225},
       adsurl = {https://ui.adsabs.harvard.edu/abs/1973ApJ...182L..85K}
}

@ARTICLE{shi2022,
       author = {{Shi}, Haoli. and {He}, Jiang. and {Liu}, Jiangtao.},
        title = "{Design of GRM Detector Simulator of SVOM Satellite}",
      journal = {Nuclear Electronics \& Detection Technology},
         year = 2022,
        month = jan,
       volume = {42},
        pages = {62},
        doi={10.3969/j.issn.0258-0934.2022.01.011}
}

@ARTICLE{abbott2017,
       author = {{Abbott}, B.~P. and {Abbott}, R. and {Abbott}, T.~D. and {Acernese}, F. and {Ackley}, K. and {Adams}, C. and {Adams}, T. and {Addesso}, P. and {Adhikari}, R.~X. and {Adya}, V.~B. and {Affeldt}, C. and {Afrough}, M. and {Agarwal}, B. and {Agathos}, M. and {Agatsuma}, K. and {Aggarwal}, N. and {Aguiar}, O.~D. and {Aiello}, L. and {Ain}, A. and {Ajith}, P. and {Allen}, B. and {Allen}, G. and {Allocca}, A. and {Aloy}, M.~A. and {Altin}, P.~A. and {Amato}, A. and {Ananyeva}, A. and {Anderson}, S.~B. and {Anderson}, W.~G. and {Angelova}, S.~V. and {Antier}, S. and {Appert}, S. and {Arai}, K. and {Araya}, M.~C. and {Areeda}, J.~S. and {Arnaud}, N. and {Arun}, K.~G. and {Ascenzi}, S. and {Ashton}, G. and {Ast}, M. and {Aston}, S.~M. and {Astone}, P. and {Atallah}, D.~V. and {Aufmuth}, P. and {Aulbert}, C. and {AultONeal}, K. and {Austin}, C. and {Avila-Alvarez}, A. and {Babak}, S. and {Bacon}, P. and {Bader}, M.~K.~M. and {Bae}, S. and {Baker}, P.~T. and {Baldaccini}, F. and {Ballardin}, G. and {Ballmer}, S.~W. and {Banagiri}, S. and {Barayoga}, J.~C. and {Barclay}, S.~E. and {Barish}, B.~C. and {Barker}, D. and {Barkett}, K. and {Barone}, F. and {Barr}, B. and {Barsotti}, L. and {Barsuglia}, M. and {Barta}, D. and {Bartlett}, J. and {Bartos}, I. and {Bassiri}, R. and {Basti}, A. and {Batch}, J.~C. and {Bawaj}, M. and {Bayley}, J.~C. and {Bazzan}, M. and {B{\'e}csy}, B. and {Beer}, C. and {Bejger}, M. and {Belahcene}, I. and {Bell}, A.~S. and {Berger}, B.~K. and {Bergmann}, G. and {Bero}, J.~J. and {Berry}, C.~P.~L. and {Bersanetti}, D. and {Bertolini}, A. and {Betzwieser}, J. and {Bhagwat}, S. and {Bhandare}, R. and {Bilenko}, I.~A. and {Billingsley}, G. and {Billman}, C.~R. and {Birch}, J. and {Birney}, R. and {Birnholtz}, O. and {Biscans}, S. and {Biscoveanu}, S. and {Bisht}, A. and {Bitossi}, M. and {Biwer}, C. and {Bizouard}, M.~A. and {Blackburn}, J.~K. and {Blackman}, J. and {Blair}, C.~D. and {Blair}, D.~G. and {Blair}, R.~M. and {Bloemen}, S. and {Bock}, O. and {Bode}, N. and {Boer}, M. and {Bogaert}, G. and {Bohe}, A. and {Bondu}, F. and {Bonilla}, E. and {Bonnand}, R. and {Boom}, B.~A. and {Bork}, R. and {Boschi}, V. and {Bose}, S. and {Bossie}, K. and {Bouffanais}, Y. and {Bozzi}, A. and {Bradaschia}, C. and {Brady}, P.~R. and {Branchesi}, M. and {Brau}, J.~E. and {Briant}, T. and {Brillet}, A. and {Brinkmann}, M. and {Brisson}, V. and {Brockill}, P. and {Broida}, J.~E. and {Brooks}, A.~F. and {Brown}, D.~A. and {Brown}, D.~D. and {Brunett}, S. and {Buchanan}, C.~C. and {Buikema}, A. and {Bulik}, T. and {Bulten}, H.~J. and {Buonanno}, A. and {Buskulic}, D. and {Buy}, C. and {Byer}, R.~L. and {Cabero}, M. and {Cadonati}, L. and {Cagnoli}, G. and {Cahillane}, C. and {Calder{\'o}n Bustillo}, J. and {Callister}, T.~A. and {Calloni}, E. and {Camp}, J.~B. and {Canepa}, M. and {Canizares}, P. and {Cannon}, K.~C. and {Cao}, H. and {Cao}, J. and {Capano}, C.~D. and {Capocasa}, E. and {Carbognani}, F. and {Caride}, S. and {Carney}, M.~F. and {Casanueva Diaz}, J. and {Casentini}, C. and {Caudill}, S. and {Cavagli{\`a}}, M. and {Cavalier}, F. and {Cavalieri}, R. and {Cella}, G. and {Cepeda}, C.~B. and {Cerd{\'a}-Dur{\'a}n}, P. and {Cerretani}, G. and {Cesarini}, E. and {Chamberlin}, S.~J. and {Chan}, M. and {Chao}, S. and {Charlton}, P. and {Chase}, E. and {Chassande-Mottin}, E. and {Chatterjee}, D. and {Chatziioannou}, K. and {Cheeseboro}, B.~D. and {Chen}, H.~Y. and {Chen}, X. and {Chen}, Y. and {Cheng}, H. -P. and {Chia}, H. and {Chincarini}, A. and {Chiummo}, A. and {Chmiel}, T. and {Cho}, H.~S. and {Cho}, M. and {Chow}, J.~H. and {Christensen}, N. and {Chu}, Q. and {Chua}, A.~J.~K. and {Chua}, S. and {Chung}, A.~K.~W. and {Chung}, S. and {Ciani}, G. and {Ciolfi}, R. and {Cirelli}, C.~E. and {Cirone}, A. and {Clara}, F. and {Clark}, J.~A. and {Clearwater}, P. and {Cleva}, F. and {Cocchieri}, C. and {Coccia}, E. and {Cohadon}, P. -F. and {Cohen}, D. and {Colla}, A. and {Collette}, C.~G. and {Cominsky}, L.~R. and {Constancio}, M., Jr. and {Conti}, L. and {Cooper}, S.~J. and {Corban}, P. and {Corbitt}, T.~R. and {Cordero-Carri{\'o}n}, I. and {Corley}, K.~R. and {Cornish}, N. and {Corsi}, A. and {Cortese}, S. and {Costa}, C.~A. and {Coughlin}, M.~W. and {Coughlin}, S.~B. and {Coulon}, J. -P. and {Countryman}, S.~T. and {Couvares}, P. and {Covas}, P.~B. and {Cowan}, E.~E. and {Coward}, D.~M. and {Cowart}, M.~J. and {Coyne}, D.~C. and {Coyne}, R. and {Creighton}, J.~D.~E. and {Creighton}, T.~D. and {Cripe}, J. and {Crowder}, S.~G. and {Cullen}, T.~J. and {Cumming}, A. and {Cunningham}, L. and {Cuoco}, E. and {Dal Canton}, T. and {D{\'a}lya}, G. and {Danilishin}, S.~L. and {D'Antonio}, S. and {Danzmann}, K. and {Dasgupta}, A. and {Da Silva Costa}, C.~F. and {Dattilo}, V. and {Dave}, I. and {Davier}, M. and {Davis}, D. and {Daw}, E.~J. and {Day}, B. and {De}, S. and {DeBra}, D. and {Degallaix}, J. and {De Laurentis}, M. and {Del{\'e}glise}, S. and {Del Pozzo}, W. and {Demos}, N. and {Denker}, T. and {Dent}, T. and {De Pietri}, R. and {Dergachev}, V. and {De Rosa}, R. and {DeRosa}, R.~T. and {De Rossi}, C. and {DeSalvo}, R. and {de Varona}, O. and {Devenson}, J. and {Dhurandhar}, S. and {D{\'\i}az}, M.~C. and {Di Fiore}, L. and {Di Giovanni}, M. and {Di Girolamo}, T. and {Di Lieto}, A. and {Di Pace}, S. and {Di Palma}, I. and {Di Renzo}, F. and {Doctor}, Z. and {Dolique}, V. and {Donovan}, F. and {Dooley}, K.~L. and {Doravari}, S. and {Dorrington}, I. and {Douglas}, R. and {Dovale {\'A}lvarez}, M. and {Downes}, T.~P. and {Drago}, M. and {Dreissigacker}, C. and {Driggers}, J.~C. and {Du}, Z. and {Ducrot}, M. and {Dupej}, P. and {Dwyer}, S.~E. and {Edo}, T.~B. and {Edwards}, M.~C. and {Effler}, A. and {Eggenstein}, H. -B. and {Ehrens}, P. and {Eichholz}, J. and {Eikenberry}, S.~S. and {Eisenstein}, R.~A. and {Essick}, R.~C. and {Estevez}, D. and {Etienne}, Z.~B. and {Etzel}, T. and {Evans}, M. and {Evans}, T.~M. and {Factourovich}, M. and {Fafone}, V. and {Fair}, H. and {Fairhurst}, S. and {Fan}, X. and {Farinon}, S. and {Farr}, B. and {Farr}, W.~M. and {Fauchon-Jones}, E.~J. and {Favata}, M. and {Fays}, M. and {Fee}, C. and {Fehrmann}, H. and {Feicht}, J. and {Fejer}, M.~M. and {Fernandez-Galiana}, A. and {Ferrante}, I. and {Ferreira}, E.~C. and {Ferrini}, F. and {Fidecaro}, F. and {Finstad}, D. and {Fiori}, I. and {Fiorucci}, D. and {Fishbach}, M. and {Fisher}, R.~P. and {Fitz-Axen}, M. and {Flaminio}, R. and {Fletcher}, M. and {Fong}, H. and {Font}, J.~A. and {Forsyth}, P.~W.~F. and {Forsyth}, S.~S. and {Fournier}, J. -D. and {Frasca}, S. and {Frasconi}, F. and {Frei}, Z. and {Freise}, A. and {Frey}, R. and {Frey}, V. and {Fries}, E.~M. and {Fritschel}, P. and {Frolov}, V.~V. and {Fulda}, P. and {Fyffe}, M. and {Gabbard}, H. and {Gadre}, B.~U. and {Gaebel}, S.~M. and {Gair}, J.~R. and {Gammaitoni}, L. and {Ganija}, M.~R. and {Gaonkar}, S.~G. and {Garcia-Quiros}, C. and {Garufi}, F. and {Gateley}, B. and {Gaudio}, S. and {Gaur}, G. and {Gayathri}, V. and {Gehrels}, N. and {Gemme}, G. and {Genin}, E. and {Gennai}, A. and {George}, D. and {George}, J. and {Gergely}, L. and {Germain}, V. and {Ghonge}, S. and {Ghosh}, Abhirup and {Ghosh}, Archisman and {Ghosh}, S. and {Giaime}, J.~A. and {Giardina}, K.~D. and {Giazotto}, A. and {Gill}, K. and {Glover}, L. and {Goetz}, E. and {Goetz}, R. and {Gomes}, S. and {Goncharov}, B. and {Gonz{\'a}lez}, G. and {Gonzalez Castro}, J.~M. and {Gopakumar}, A. and {Gorodetsky}, M.~L. and {Gossan}, S.~E. and {Gosselin}, M. and {Gouaty}, R. and {Grado}, A. and {Graef}, C. and {Granata}, M. and {Grant}, A. and {Gras}, S. and {Gray}, C. and {Greco}, G. and {Green}, A.~C. and {Gretarsson}, E.~M. and {Groot}, P. and {Grote}, H. and {Grunewald}, S. and {Gruning}, P. and {Guidi}, G.~M. and {Guo}, X. and {Gupta}, A. and {Gupta}, M.~K. and {Gushwa}, K.~E. and {Gustafson}, E.~K. and {Gustafson}, R. and {Halim}, O. and {Hall}, B.~R. and {Hall}, E.~D. and {Hamilton}, E.~Z. and {Hammond}, G. and {Haney}, M. and {Hanke}, M.~M. and {Hanks}, J. and {Hanna}, C. and {Hannam}, M.~D. and {Hannuksela}, O.~A. and {Hanson}, J. and {Hardwick}, T. and {Harms}, J. and {Harry}, G.~M. and {Harry}, I.~W. and {Hart}, M.~J. and {Haster}, C. -J. and {Haughian}, K. and {Healy}, J. and {Heidmann}, A. and {Heintze}, M.~C. and {Heitmann}, H. and {Hello}, P. and {Hemming}, G. and {Hendry}, M. and {Heng}, I.~S. and {Hennig}, J. and {Heptonstall}, A.~W. and {Heurs}, M. and {Hild}, S. and {Hinderer}, T. and {Hoak}, D. and {Hofman}, D. and {Holt}, K. and {Holz}, D.~E. and {Hopkins}, P. and {Horst}, C. and {Hough}, J. and {Houston}, E.~A. and {Howell}, E.~J. and {Hreibi}, A. and {Hu}, Y.~M. and {Huerta}, E.~A. and {Huet}, D. and {Hughey}, B. and {Husa}, S. and {Huttner}, S.~H. and {Huynh-Dinh}, T. and {Indik}, N. and {Inta}, R. and {Intini}, G. and {Isa}, H.~N. and {Isac}, J. -M. and {Isi}, M. and {Iyer}, B.~R. and {Izumi}, K. and {Jacqmin}, T. and {Jani}, K. and {Jaranowski}, P. and {Jawahar}, S. and {Jim{\'e}nez-Forteza}, F. and {Johnson}, W.~W. and {Johnson-McDaniel}, N.~K. and {Jones}, D.~I. and {Jones}, R. and {Jonker}, R.~J.~G. and {Ju}, L. and {Junker}, J. and {Kalaghatgi}, C.~V. and {Kalogera}, V. and {Kamai}, B. and {Kandhasamy}, S. and {Kang}, G. and {Kanner}, J.~B. and {Kapadia}, S.~J. and {Karki}, S. and {Karvinen}, K.~S. and {Kasprzack}, M. and {Kastaun}, W. and {Katolik}, M. and {Katsavounidis}, E. and {Katzman}, W. and {Kaufer}, S. and {Kawabe}, K. and {K{\'e}f{\'e}lian}, F. and {Keitel}, D. and {Kemball}, A.~J. and {Kennedy}, R. and {Kent}, C. and {Key}, J.~S. and {Khalili}, F.~Y. and {Khan}, I. and {Khan}, S. and {Khan}, Z. and {Khazanov}, E.~A. and {Kijbunchoo}, N. and {Kim}, Chunglee and {Kim}, J.~C. and {Kim}, K. and {Kim}, W. and {Kim}, W.~S. and {Kim}, Y. -M. and {Kimbrell}, S.~J. and {King}, E.~J. and {King}, P.~J. and {Kinley-Hanlon}, M. and {Kirchhoff}, R. and {Kissel}, J.~S. and {Kleybolte}, L. and {Klimenko}, S. and {Knowles}, T.~D. and {Koch}, P. and {Koehlenbeck}, S.~M. and {Koley}, S. and {Kondrashov}, V. and {Kontos}, A. and {Korobko}, M. and {Korth}, W.~Z. and {Kowalska}, I. and {Kozak}, D.~B. and {Kr{\"a}mer}, C. and {Kringel}, V. and {Krishnan}, B. and {Kr{\'o}lak}, A. and {Kuehn}, G. and {Kumar}, P. and {Kumar}, R. and {Kumar}, S. and {Kuo}, L. and {Kutynia}, A. and {Kwang}, S. and {Lackey}, B.~D. and {Lai}, K.~H. and {Landry}, M. and {Lang}, R.~N. and {Lange}, J. and {Lantz}, B. and {Lanza}, R.~K. and {Lartaux-Vollard}, A. and {Lasky}, P.~D. and {Laxen}, M. and {Lazzarini}, A. and {Lazzaro}, C. and {Leaci}, P. and {Leavey}, S. and {Lee}, C.~H. and {Lee}, H.~K. and {Lee}, H.~M. and {Lee}, H.~W. and {Lee}, K. and {Lehmann}, J. and {Lenon}, A. and {Leonardi}, M. and {Leroy}, N. and {Letendre}, N. and {Levin}, Y. and {Li}, T.~G.~F. and {Linker}, S.~D. and {Littenberg}, T.~B. and {Liu}, J. and {Lo}, R.~K.~L. and {Lockerbie}, N.~A. and {London}, L.~T. and {Lord}, J.~E. and {Lorenzini}, M. and {Loriette}, V. and {Lormand}, M. and {Losurdo}, G. and {Lough}, J.~D. and {Lousto}, C.~O. and {Lovelace}, G. and {L{\"u}ck}, H. and {Lumaca}, D. and {Lundgren}, A.~P. and {Lynch}, R. and {Ma}, Y. and {Macas}, R. and {Macfoy}, S. and {Machenschalk}, B. and {MacInnis}, M. and {Macleod}, D.~M. and {Maga{\~n}a Hernandez}, I. and {Maga{\~n}a-Sandoval}, F. and {Maga{\~n}a Zertuche}, L. and {Magee}, R.~M. and {Majorana}, E. and {Maksimovic}, I. and {Man}, N. and {Mandic}, V. and {Mangano}, V. and {Mansell}, G.~L. and {Manske}, M. and {Mantovani}, M. and {Marchesoni}, F. and {Marion}, F. and {M{\'a}rka}, S. and {M{\'a}rka}, Z. and {Markakis}, C. and {Markosyan}, A.~S. and {Markowitz}, A. and {Maros}, E. and {Marquina}, A. and {Martelli}, F. and {Martellini}, L. and {Martin}, I.~W. and {Martin}, R.~M. and {Martynov}, D.~V. and {Mason}, K. and {Massera}, E. and {Masserot}, A. and {Massinger}, T.~J. and {Masso-Reid}, M. and {Mastrogiovanni}, S. and {Matas}, A. and {Matichard}, F. and {Matone}, L. and {Mavalvala}, N. and {Mazumder}, N. and {McCarthy}, R. and {McClelland}, D.~E. and {McCormick}, S. and {McCuller}, L. and {McGuire}, S.~C. and {McIntyre}, G. and {McIver}, J. and {McManus}, D.~J. and {McNeill}, L. and {McRae}, T. and {McWilliams}, S.~T. and {Meacher}, D. and {Meadors}, G.~D. and {Mehmet}, M. and {Meidam}, J. and {Mejuto-Villa}, E. and {Melatos}, A. and {Mendell}, G. and {Mercer}, R.~A. and {Merilh}, E.~L. and {Merzougui}, M. and {Meshkov}, S. and {Messenger}, C. and {Messick}, C. and {Metzdorff}, R. and {Meyers}, P.~M. and {Miao}, H. and {Michel}, C. and {Middleton}, H. and {Mikhailov}, E.~E. and {Milano}, L. and {Miller}, A.~L. and {Miller}, B.~B. and {Miller}, J. and {Millhouse}, M. and {Milovich-Goff}, M.~C. and {Minazzoli}, O. and {Minenkov}, Y. and {Ming}, J. and {Mishra}, C. and {Mitra}, S. and {Mitrofanov}, V.~P. and {Mitselmakher}, G. and {Mittleman}, R. and {Moffa}, D. and {Moggi}, A. and {Mogushi}, K. and {Mohan}, M. and {Mohapatra}, S.~R.~P. and {Montani}, M. and {Moore}, C.~J. and {Moraru}, D. and {Moreno}, G. and {Morriss}, S.~R. and {Mours}, B. and {Mow-Lowry}, C.~M. and {Mueller}, G. and {Muir}, A.~W. and {Mukherjee}, Arunava and {Mukherjee}, D. and {Mukherjee}, S. and {Mukund}, N. and {Mullavey}, A. and {Munch}, J. and {Mu{\~n}iz}, E.~A. and {Muratore}, M. and {Murray}, P.~G. and {Napier}, K. and {Nardecchia}, I. and {Naticchioni}, L. and {Nayak}, R.~K. and {Neilson}, J. and {Nelemans}, G. and {Nelson}, T.~J.~N. and {Nery}, M. and {Neunzert}, A. and {Nevin}, L. and {Newport}, J.~M. and {Newton}, G. and {Ng}, K.~K.~Y. and {Nguyen}, T.~T. and {Nichols}, D. and {Nielsen}, A.~B. and {Nissanke}, S. and {Nitz}, A. and {Noack}, A. and {Nocera}, F. and {Nolting}, D. and {North}, C. and {Nuttall}, L.~K. and {Oberling}, J. and {O'Dea}, G.~D. and {Ogin}, G.~H. and {Oh}, J.~J. and {Oh}, S.~H. and {Ohme}, F. and {Okada}, M.~A. and {Oliver}, M. and {Oppermann}, P. and {Oram}, Richard J. and {O'Reilly}, B. and {Ormiston}, R. and {Ortega}, L.~F. and {O'Shaughnessy}, R. and {Ossokine}, S. and {Ottaway}, D.~J. and {Overmier}, H. and {Owen}, B.~J. and {Pace}, A.~E. and {Page}, J. and {Page}, M.~A. and {Pai}, A. and {Pai}, S.~A. and {Palamos}, J.~R. and {Palashov}, O. and {Palomba}, C. and {Pal-Singh}, A. and {Pan}, Howard and {Pan}, Huang-Wei and {Pang}, B. and {Pang}, P.~T.~H. and {Pankow}, C. and {Pannarale}, F. and {Pant}, B.~C. and {Paoletti}, F. and {Paoli}, A. and {Papa}, M.~A. and {Parida}, A. and {Parker}, W. and {Pascucci}, D. and {Pasqualetti}, A. and {Passaquieti}, R. and {Passuello}, D. and {Patil}, M. and {Patricelli}, B. and {Pearlstone}, B.~L. and {Pedraza}, M. and {Pedurand}, R. and {Pekowsky}, L. and {Pele}, A. and {Penn}, S. and {Perez}, C.~J. and {Perreca}, A. and {Perri}, L.~M. and {Pfeiffer}, H.~P. and {Phelps}, M. and {Piccinni}, O.~J. and {Pichot}, M. and {Piergiovanni}, F. and {Pierro}, V. and {Pillant}, G. and {Pinard}, L. and {Pinto}, I.~M. and {Pirello}, M. and {Pitkin}, M. and {Poe}, M. and {Poggiani}, R. and {Popolizio}, P. and {Porter}, E.~K. and {Post}, A. and {Powell}, J. and {Prasad}, J. and {Pratt}, J.~W.~W. and {Pratten}, G. and {Predoi}, V. and {Prestegard}, T. and {Prijatelj}, M. and {Principe}, M. and {Privitera}, S. and {Prodi}, G.~A. and {Prokhorov}, L.~G. and {Puncken}, O. and {Punturo}, M. and {Puppo}, P. and {P{\"u}rrer}, M. and {Qi}, H. and {Quetschke}, V. and {Quintero}, E.~A. and {Quitzow-James}, R. and {Raab}, F.~J. and {Rabeling}, D.~S. and {Radkins}, H. and {Raffai}, P. and {Raja}, S. and {Rajan}, C. and {Rajbhandari}, B. and {Rakhmanov}, M. and {Ramirez}, K.~E. and {Ramos-Buades}, A. and {Rapagnani}, P. and {Raymond}, V. and {Razzano}, M. and {Read}, J. and {Regimbau}, T. and {Rei}, L. and {Reid}, S. and {Reitze}, D.~H. and {Ren}, W. and {Reyes}, S.~D. and {Ricci}, F. and {Ricker}, P.~M. and {Rieger}, S. and {Riles}, K. and {Rizzo}, M. and {Robertson}, N.~A. and {Robie}, R. and {Robinet}, F. and {Rocchi}, A. and {Rolland}, L. and {Rollins}, J.~G. and {Roma}, V.~J. and {Romano}, R. and {Romel}, C.~L. and {Romie}, J.~H. and {Rosi{\'n}ska}, D. and {Ross}, M.~P. and {Rowan}, S. and {R{\"u}diger}, A. and {Ruggi}, P. and {Rutins}, G. and {Ryan}, K. and {Sachdev}, S. and {Sadecki}, T. and {Sadeghian}, L. and {Sakellariadou}, M. and {Salconi}, L. and {Saleem}, M. and {Salemi}, F. and {Samajdar}, A. and {Sammut}, L. and {Sampson}, L.~M. and {Sanchez}, E.~J. and {Sanchez}, L.~E. and {Sanchis-Gual}, N. and {Sandberg}, V. and {Sanders}, J.~R. and {Sassolas}, B. and {Sathyaprakash}, B.~S. and {Saulson}, P.~R. and {Sauter}, O. and {Savage}, R.~L. and {Sawadsky}, A. and {Schale}, P. and {Scheel}, M. and {Scheuer}, J. and {Schmidt}, J. and {Schmidt}, P. and {Schnabel}, R. and {Schofield}, R.~M.~S. and {Sch{\"o}nbeck}, A. and {Schreiber}, E. and {Schuette}, D. and {Schulte}, B.~W. and {Schutz}, B.~F. and {Schwalbe}, S.~G. and {Scott}, J. and {Scott}, S.~M. and {Seidel}, E. and {Sellers}, D. and {Sengupta}, A.~S. and {Sentenac}, D. and {Sequino}, V. and {Sergeev}, A. and {Shaddock}, D.~A. and {Shaffer}, T.~J. and {Shah}, A.~A. and {Shahriar}, M.~S. and {Shaner}, M.~B. and {Shao}, L. and {Shapiro}, B. and {Shawhan}, P. and {Sheperd}, A. and {Shoemaker}, D.~H. and {Shoemaker}, D.~M. and {Siellez}, K. and {Siemens}, X. and {Sieniawska}, M. and {Sigg}, D. and {Silva}, A.~D. and {Singer}, L.~P. and {Singh}, A. and {Singhal}, A. and {Sintes}, A.~M. and {Slagmolen}, B.~J.~J. and {Smith}, B. and {Smith}, J.~R. and {Smith}, R.~J.~E. and {Somala}, S. and {Son}, E.~J. and {Sonnenberg}, J.~A. and {Sorazu}, B. and {Sorrentino}, F. and {Souradeep}, T. and {Spencer}, A.~P. and {Srivastava}, A.~K. and {Staats}, K. and {Staley}, A. and {Steinke}, M. and {Steinlechner}, J. and {Steinlechner}, S. and {Steinmeyer}, D. and {Stevenson}, S.~P. and {Stone}, R. and {Stops}, D.~J. and {Strain}, K.~A. and {Stratta}, G. and {Strigin}, S.~E. and {Strunk}, A. and {Sturani}, R. and {Stuver}, A.~L. and {Summerscales}, T.~Z. and {Sun}, L. and {Sunil}, S. and {Suresh}, J. and {Sutton}, P.~J. and {Swinkels}, B.~L. and {Szczepa{\'n}czyk}, M.~J. and {Tacca}, M. and {Tait}, S.~C. and {Talbot}, C. and {Talukder}, D. and {Tanner}, D.~B. and {T{\'a}pai}, M. and {Taracchini}, A. and {Tasson}, J.~D. and {Taylor}, J.~A. and {Taylor}, R. and {Tewari}, S.~V. and {Theeg}, T. and {Thies}, F. and {Thomas}, E.~G. and {Thomas}, M. and {Thomas}, P. and {Thorne}, K.~A. and {Thorne}, K.~S. and {Thrane}, E. and {Tiwari}, S. and {Tiwari}, V. and {Tokmakov}, K.~V. and {Toland}, K. and {Tonelli}, M. and {Tornasi}, Z. and {Torres-Forn{\'e}}, A. and {Torrie}, C.~I. and {T{\"o}yr{\"a}}, D. and {Travasso}, F. and {Traylor}, G. and {Trinastic}, J. and {Tringali}, M.~C. and {Trozzo}, L. and {Tsang}, K.~W. and {Tse}, M. and {Tso}, R. and {Tsukada}, L. and {Tsuna}, D. and {Tuyenbayev}, D. and {Ueno}, K. and {Ugolini}, D. and {Unnikrishnan}, C.~S. and {Urban}, A.~L. and {Usman}, S.~A. and {Vahlbruch}, H. and {Vajente}, G. and {Valdes}, G. and {van Bakel}, N. and {van Beuzekom}, M. and {van den Brand}, J.~F.~J. and {Van Den Broeck}, C. and {Vander-Hyde}, D.~C. and {van der Schaaf}, L. and {van Heijningen}, J.~V. and {van Veggel}, A.~A. and {Vardaro}, M. and {Varma}, V. and {Vass}, S. and {Vas{\'u}th}, M. and {Vecchio}, A. and {Vedovato}, G. and {Veitch}, J. and {Veitch}, P.~J. and {Venkateswara}, K. and {Venugopalan}, G. and {Verkindt}, D. and {Vetrano}, F. and {Vicer{\'e}}, A. and {Viets}, A.~D. and {Vinciguerra}, S. and {Vine}, D.~J. and {Vinet}, J. -Y. and {Vitale}, S. and {Vo}, T. and {Vocca}, H. and {Vorvick}, C. and {Vyatchanin}, S.~P. and {Wade}, A.~R. and {Wade}, L.~E. and {Wade}, M. and {Walet}, R. and {Walker}, M. and {Wallace}, L. and {Walsh}, S. and {Wang}, G. and {Wang}, H. and {Wang}, J.~Z. and {Wang}, W.~H. and {Wang}, Y.~F. and {Ward}, R.~L. and {Warner}, J. and {Was}, M. and {Watchi}, J. and {Weaver}, B. and {Wei}, L. -W. and {Weinert}, M. and {Weinstein}, A.~J. and {Weiss}, R. and {Wen}, L. and {Wessel}, E.~K. and {We{\ss}els}, P. and {Westerweck}, J. and {Westphal}, T. and {Wette}, K. and {Whelan}, J.~T. and {Whitcomb}, S.~E. and {Whiting}, B.~F. and {Whittle}, C. and {Wilken}, D. and {Williams}, D. and {Williams}, R.~D. and {Williamson}, A.~R. and {Willis}, J.~L. and {Willke}, B. and {Wimmer}, M.~H. and {Winkler}, W. and {Wipf}, C.~C. and {Wittel}, H. and {Woan}, G. and {Woehler}, J. and {Wofford}, J. and {Wong}, K.~W.~K. and {Worden}, J. and {Wright}, J.~L. and {Wu}, D.~S. and {Wysocki}, D.~M. and {Xiao}, S. and {Yamamoto}, H. and {Yancey}, C.~C. and {Yang}, L. and {Yap}, M.~J. and {Yazback}, M. and {Yu}, Hang and {Yu}, Haocun and {Yvert}, M. and {Zadro{\.z}ny}, A. and {Zanolin}, M. and {Zelenova}, T. and {Zendri}, J. -P. and {Zevin}, M. and {Zhang}, L. and {Zhang}, M. and {Zhang}, T. and {Zhang}, Y. -H. and {Zhao}, C. and {Zhou}, M. and {Zhou}, Z. and {Zhu}, S.~J. and {Zhu}, X.~J. and {Zimmerman}, A.~B. and {Zucker}, M.~E. and {Zweizig}, J. and {(LIGO Scientific Collaboration} and {Virgo Collaboration} and {Burns}, E. and {Veres}, P. and {Kocevski}, D. and {Racusin}, J. and {Goldstein}, A. and {Connaughton}, V. and {Briggs}, M.~S. and {Blackburn}, L. and {Hamburg}, R. and {Hui}, C.~M. and {von Kienlin}, A. and {McEnery}, J. and {Preece}, R.~D. and {Wilson-Hodge}, C.~A. and {Bissaldi}, E. and {Cleveland}, W.~H. and {Gibby}, M.~H. and {Giles}, M.~M. and {Kippen}, R.~M. and {McBreen}, S. and {Meegan}, C.~A. and {Paciesas}, W.~S. and {Poolakkil}, S. and {Roberts}, O.~J. and {Stanbro}, M. and {Gamma-ray Burst Monitor}, (Fermi and {Savchenko}, V. and {Ferrigno}, C. and {Kuulkers}, E. and {Bazzano}, A. and {Bozzo}, E. and {Brandt}, S. and {Chenevez}, J. and {Courvoisier}, T.~J. -L. and {Diehl}, R. and {Domingo}, A. and {Hanlon}, L. and {Jourdain}, E. and {Laurent}, P. and {Lebrun}, F. and {Lutovinov}, A. and {Mereghetti}, S. and {Natalucci}, L. and {Rodi}, J. and {Roques}, J. -P. and {Sunyaev}, R. and {Ubertini}, P. and {(INTEGRAL}},
        title = "{Gravitational Waves and Gamma-Rays from a Binary Neutron Star Merger: GW170817 and GRB 170817A}",
      journal = {\apjl},
         year = 2017,
        month = oct,
       volume = {848},
       number = {2},
          eid = {L13},
        pages = {L13},
          doi = {10.3847/2041-8213/aa920c},
archivePrefix = {arXiv},
       eprint = {1710.05834},
 primaryClass = {astro-ph.HE},
       adsurl = {https://ui.adsabs.harvard.edu/abs/2017ApJ...848L..13A}
}

@ARTICLE{li2018,
       author = {{Li}, TiPei and {Xiong}, ShaoLin and {Zhang}, ShuangNan and {Lu}, FangJun and {Song}, LiMing and {Cao}, XueLei and {Chang}, Zhi and {Chen}, Gang and {Chen}, Li and {Chen}, TianXiang and {Chen}, Yong and {Chen}, YiBao and {Chen}, YuPeng and {Cui}, Wei and {Cui}, WeiWei and {Deng}, JingKang and {Dong}, YongWei and {Du}, YuanYuan and {Fu}, MinXue and {Gao}, GuanHua and {Gao}, He and {Gao}, Min and {Ge}, MingYu and {Gu}, YuDong and {Guan}, Ju and {Guo}, ChengCheng and {Han}, DaWei and {Hu}, Wei and {Huang}, Yue and {Huo}, Jia and {Jia}, ShuMei and {Jiang}, LuHua and {Jiang}, WeiChun and {Jin}, Jing and {Jin}, YongJie and {Li}, Bing and {Li}, ChengKui and {Li}, Gang and {Li}, MaoShun and {Li}, Wei and {Li}, Xian and {Li}, XiaoBo and {Li}, XuFang and {Li}, YanGuo and {Li}, ZiJian and {Li}, ZhengWei and {Liang}, XiaoHua and {Liao}, JinYuan and {Liu}, CongZhan and {Liu}, GuoQing and {Liu}, HongWei and {Liu}, ShaoZhen and {Liu}, XiaoJing and {Liu}, Yuan and {Liu}, YiNong and {Lu}, Bo and {Lu}, XueFeng and {Luo}, Tao and {Ma}, Xiang and {Meng}, Bin and {Nang}, Yi and {Nie}, JianYin and {Ou}, Ge and {Qu}, JinLu and {Sai}, Na and {Sun}, Liang and {Tan}, Yin and {Tao}, Lian and {Tao}, WenHui and {Tuo}, YouLi and {Wang}, GuoFeng and {Wang}, HuanYu and {Wang}, Juan and {Wang}, WenShuai and {Wang}, YuSa and {Wen}, XiangYang and {Wu}, BoBing and {Wu}, Mei and {Xiao}, GuangCheng and {Xu}, He and {Xu}, YuPeng and {Yan}, LinLi and {Yang}, JiaWei and {Yang}, Sheng and {Yang}, YanJi and {Zhang}, AiMei and {Zhang}, ChunLei and {Zhang}, ChengMo and {Zhang}, Fan and {Zhang}, HongMei and {Zhang}, Juan and {Zhang}, Qiang and {Zhang}, Shu and {Zhang}, Tong and {Zhang}, Wei and {Zhang}, WanChang and {Zhang}, WenZhao and {Zhang}, Yi and {Zhang}, Yue and {Zhang}, YiFei and {Zhang}, YongJie and {Zhang}, Zhao and {Zhang}, ZiLiang and {Zhao}, HaiSheng and {Zhao}, JianLing and {Zhao}, XiaoFan and {Zheng}, ShiJie and {Zhu}, Yue and {Zhu}, YuXuan and {Zou}, ChangLin},
        title = "{Insight-HXMT observations of the first binary neutron star merger GW170817}",
      journal = {Science China Physics, Mechanics, and Astronomy},
         year = 2018,
        month = mar,
       volume = {61},
       number = {3},
          eid = {31011},
        pages = {31011},
          doi = {10.1007/s11433-017-9107-5},
archivePrefix = {arXiv},
       eprint = {1710.06065},
 primaryClass = {astro-ph.HE},
       adsurl = {https://ui.adsabs.harvard.edu/abs/2018SCPMA..61c1011L}
}

@ARTICLE{goldstein2017,
       author = {{Goldstein}, A. and {Veres}, P. and {Burns}, E. and {Briggs}, M.~S. and {Hamburg}, R. and {Kocevski}, D. and {Wilson-Hodge}, C.~A. and {Preece}, R.~D. and {Poolakkil}, S. and {Roberts}, O.~J. and {Hui}, C.~M. and {Connaughton}, V. and {Racusin}, J. and {von Kienlin}, A. and {Dal Canton}, T. and {Christensen}, N. and {Littenberg}, T. and {Siellez}, K. and {Blackburn}, L. and {Broida}, J. and {Bissaldi}, E. and {Cleveland}, W.~H. and {Gibby}, M.~H. and {Giles}, M.~M. and {Kippen}, R.~M. and {McBreen}, S. and {McEnery}, J. and {Meegan}, C.~A. and {Paciesas}, W.~S. and {Stanbro}, M.},
        title = "{An Ordinary Short Gamma-Ray Burst with Extraordinary Implications: Fermi-GBM Detection of GRB 170817A}",
      journal = {\apjl},
         year = 2017,
        month = oct,
       volume = {848},
       number = {2},
          eid = {L14},
        pages = {L14},
          doi = {10.3847/2041-8213/aa8f41},
archivePrefix = {arXiv},
       eprint = {1710.05446},
 primaryClass = {astro-ph.HE},
       adsurl = {https://ui.adsabs.harvard.edu/abs/2017ApJ...848L..14G}
}

@ARTICLE{zhang2011,
       author = {{Zhang}, Bing},
        title = "{Open questions in GRB physics}",
      journal = {Comptes Rendus Physique},
         year = 2011,
        month = apr,
       volume = {12},
        pages = {206-225},
          doi = {10.1016/j.crhy.2011.03.004},
archivePrefix = {arXiv},
       eprint = {1104.0932},
 primaryClass = {astro-ph.HE},
       adsurl = {https://ui.adsabs.harvard.edu/abs/2011CRPhy..12..206Z}
}

@BOOK{zhang2018,
       author = {{Zhang}, Bing},
        title = "{The Physics of Gamma-Ray Bursts}",
         year = 2018,
          doi = {10.1017/9781139226530},
       adsurl = {https://ui.adsabs.harvard.edu/abs/2018pgrb.book.....Z}
}

@ARTICLE{burgess2018,
       author = {{Burgess}, J. Michael and {Yu}, Hoi-Fung and {Greiner}, Jochen and {Mortlock}, Daniel J.},
        title = "{Awakening the BALROG: BAyesian Location Reconstruction Of GRBs}",
      journal = {\mnras},
         year = 2018,
        month = may,
       volume = {476},
       number = {2},
        pages = {1427-1444},
          doi = {10.1093/mnras/stx2853},
archivePrefix = {arXiv},
       eprint = {1610.07385},
 primaryClass = {astro-ph.IM},
       adsurl = {https://ui.adsabs.harvard.edu/abs/2018MNRAS.476.1427B}
}

@ARTICLE{pendleton1999,
       author = {{Pendleton}, Geoffrey N. and {Briggs}, Michael S. and {Kippen}, R. Marc and {Paciesas}, William. S. and {Stollberg}, Mark and {Woods}, Pete and {Meegan}, Charles A. and {Fishman}, Gerald J. and {McCollough}, Mike L. and {Connaughton}, Valerie},
        title = "{The Structure and Evolution of LOCBURST: The BATSE Burst Location Algorithm}",
      journal = {\apj},
         year = 1999,
        month = feb,
       volume = {512},
       number = {1},
        pages = {362-376},
          doi = {10.1086/306735},
       adsurl = {https://ui.adsabs.harvard.edu/abs/1999ApJ...512..362P}
}

@ARTICLE{connaughton2015,
       author = {{Connaughton}, V. and {Briggs}, M.~S. and {Goldstein}, A. and {Meegan}, C.~A. and {Paciesas}, W.~S. and {Preece}, R.~D. and {Wilson-Hodge}, C.~A. and {Gibby}, M.~H. and {Greiner}, J. and {Gruber}, D. and {Jenke}, P. and {Kippen}, R.~M. and {Pelassa}, V. and {Xiong}, S. and {Yu}, H. -F. and {Bhat}, P.~N. and {Burgess}, J.~M. and {Byrne}, D. and {Fitzpatrick}, G. and {Foley}, S. and {Giles}, M.~M. and {Guiriec}, S. and {van der Horst}, A.~J. and {von Kienlin}, A. and {McBreen}, S. and {McGlynn}, S. and {Tierney}, D. and {Zhang}, B. -B.},
        title = "{Localization of Gamma-Ray Bursts Using the Fermi Gamma-Ray Burst Monitor}",
      journal = {\apjs},
         year = 2015,
        month = feb,
       volume = {216},
       number = {2},
          eid = {32},
        pages = {32},
          doi = {10.1088/0067-0049/216/2/32},
archivePrefix = {arXiv},
       eprint = {1411.2685},
 primaryClass = {astro-ph.IM},
       adsurl = {https://ui.adsabs.harvard.edu/abs/2015ApJS..216...32C}
}

@ARTICLE{GCN36787,
       author = {{Fermi GBM Team}},
        title = "{GRB 240629A: BALROG localization (Fermi Trigger 741372837 / GRB 240629704)}",
      journal = {GRB Coordinates Network},
         year = 2024,
        month = jun,
       volume = {36787},
        pages = {1}
}
\end{document}